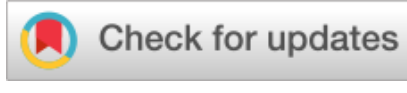

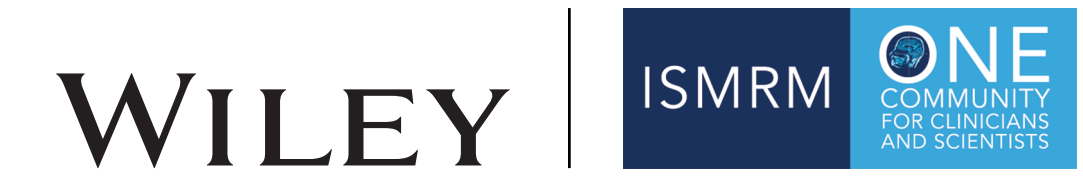

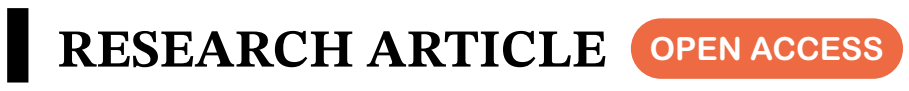

RESEARCH ARTICLE OPEN ACCESS

# Group-Patch Joint Compression: Compressing Dynamic $B_0$ and Static RF Spatial Modulations Across k-Space Subregion Groups for Highly Accelerated MRI

Rui Tian[1] | Klaus Scheffler[1,2]

[1]High-Field MR Center, Max Planck Institute for Biological Cybernetics, Tübingen, Germany | [2]Department for Biomedical Magnetic Resonance, University of Tübingen, Tübingen, Germany

**Correspondence:** Rui Tian (rui.tian@tuebingen.mpg.de)

## ABSTRACT

**Purpose:** To accelerate MRI further, rapid $B_0$ field modulations can be applied during oversampled readout to capture additional physical information, as in Wave-CAIPI/FRONSAC/local $B_0$ coils modulation techniques. These methods, however, turn the Fourier readout into a non-Fourier-encoded dimension that cannot be reconstructed by FFT, posing significant reconstruction challenges especially in compressed-sensing or neural-network frameworks.

**Theory and Methods:** Because the rapid $B_0$ modulations still vary slowly relative to the oversampled ADC dwell time, we exploit this encoding redundancy by compressing k-space patch-by-patch across subregions, each of which is jointly encoded by a distinct subset of $B_0$ and RF (receive) spatial encoding functions. For each subset, a compression matrix is computed once and reused to compress all patches encoded by the same $B_0$-RF spatial modulations. This can be implemented by feeding subsets of $B_0$ and RF spatial encoding maps into an adapted conventional RF array compression algorithm, mimicking an expanded set of virtual receiver channels. This approach was evaluated on human brain scans at 9.4 T/3 T.

**Results:** The proposed group-patch joint compression achieves substantially higher compression factors than conventional RF-only compression, while minimally compromising encoding efficiency. Typically, joint compression factors of 11×–20× led to negligible encoding loss, dramatically reducing reconstruction time and peak memory usage. For example, compressed-sensing reconstruction took 1.4–5.1 s/2D slice, 177 s–10.1 min/3D volume on a high-memory CPU node.

**Conclusion:** Given joint encoding of dynamic $B_0$ and static RF fields, compressing multidimensional k-space patches in separate groups outperforms compressing RF receivers alone. This substantially mitigates a fundamental computational bottleneck when combining rapid $B_0$ and RF-receiver modulations.

## 1 | Introduction

MRI [1] with conventional k-space [2–4] sampling is sequential and thus, inherently slow. Two basic strategies have emerged to effectively reduce scan time: capturing more physically-encoded information within a shorter time window [5–16], and reconstructing high-quality images from undersampled data using advanced numerical methods [17–25]. These can be mathematically described by the classical forward model y = Ex [10, 26, 27] with additional regularizations.

When pushing k-space information throughput to its limits by utilizing both rapid $B_0$ and RF-receive field modulations, such as in Wave-CAIPI [16]/FRONSAC [28, 29]/local $B_0$ coils modulations [30], the Fourier readout will be turned into another oversampled (e.g., 8×), non-Fourier-encoded dimension [14–16, 28, 30–32] in addition to the RF receivers dimension [9–11, 33]. This leads to a reconstruction bottleneck [34], hindering their widespread adoption. Therefore, we propose a novel MRI compression strategy, called group-patch joint compression [35–37], to compress k-space data patches encoded by different $B_0$ and RF receive ($B_1^-$) spatial encoding functions (represented by image-space maps, or k-space kernels) in separate groups, each spanning a selected k-space subregion and all receiver channels. This generalizes compressions [38–42] of RF receiver array [43–46] (i.e., RF sensitivity maps [24] or GRAPPA kernels [11, 47]) to further compress data encoded by rapid $B_0$ field modulations (i.e., dynamic $B_0$ phase evolution maps [48] or kernels [49]). Note that, for the latter, the k-space coverage is determined by the time integral of gradient modulation currents, rather than by the maximum meaningful [47] GRAPPA kernel size that is constrained by the RF wavelength [50, 51].

Specifically, to surpass the parallel imaging acceleration limits, one could employ a faster k-space trajectory using high-performance gradients [52–57], or introduce rapid local wiggles [14–16, 28, 30, 32, 55, 58] to broaden [30, 47, 59] the trajectory's coverage. Both enrich the physical information acquired by the encoding operator, and are restricted by merely practical limits [60–62] due to peripheral-nerve-stimulation (PNS) and hardware capability. Currently, while high-performance gradients are increasingly available, applying rapid $B_0$ fields modulations for trajectory wiggles remains much less accessible. Although system imperfections (e.g., eddy currents, concomitant fields) induced by stronger gradient modulations [63] can be captured by field camera [64–66] or $B_0$ kernels' auto-calibration [67], a remaining fundamental challenge is that, fully capturing the spin signals modulated by rapid $B_0$ fields demands comparably fast ADC oversampling. This considerably increases the encoding matrix size along non-Fourier encoded dimensions. Although the Siemens ICE environment with parallel CPUs can accelerate the least-squares reconstruction for 1 $mm^3$ volumetric Wave-CAIPI scans to within a few minutes [41, 68, 69], higher resolution acquisitions and the compressed-sensing (CS) or neural-network-assisted iterative reconstructions remain critically constrained by computational burden [34]. Namely, the scan time becomes shorter, but consequently, the reconstruction time becomes longer.

Despite growing computational power, aggressively modulated and oversampled scans demand commensurately aggressive data compression to avoid prohibitive reconstruction time. Unfortunately, using brute-force SVD operation [70] to compress the entire k-space in a single step is impractical. Here, a key insight is that, the growth factor of the non-Fourier encoding dimensions (e.g., 8× oversampling, 32 RF receivers) usually far exceeds reasonable acceleration factors (e.g., 4 × 3 for human brain), leading to a substantial increase in data volume despite the substantially reduced acquisition time. This mismatch indicates significant encoding redundancy among neighboring readout samples along wiggled trajectories, since the additionally superimposed rapid $B_0$ fields still vary slowly (typically 2–20 kHz, sinusoidal) relative to the oversampled ADC dwell time (e.g., 3 $\mu s$).

Thus, the proposed group-patch joint compression technique compresses the entire k-space patch-by-patch, based on predefined k-space subregion groups. Its novelty lies in leveraging the periodicity of the rapid $B_0$ field modulations, which is typically sinusoidal along readout and identical across TRs. Specifically, because the $B_0$ modulation pattern repeats, we compute a set of compression matrices for small-size k-space patches (e.g., each with 9 readout points, 32 receivers) whose samples fall within specific intervals of a $B_0$ modulation cycle, corresponding to groups spanning different adjacent intervals within one sinusoidal period. These group-specific compression matrices are computed once and reused to compress all k-space patches assigned to the same group indices that share identical $B_0$-RF encoding functions across different k-space subregions and phase-encoding steps. Consequently, the full k-space compression reduces to patch-wise compression using only a small set of reusable group-specific compression matrices. To compute a compression matrix for a subregion group, we adapt a conventional RF array compression algorithm [38], by feeding it with all spatial encoding maps (instead of RF receiver maps only) induced by scanner gradients, local $B_0$ coils, and multiple RF receivers that encode this k-space subregion.

To characterize and experimentally test our method, we perform 2D and 3D Cartesian scans accelerated by the local $B_0$ coils modulations [30] (9.4 T) or by the scanner's linear gradient modulations (both 9.4 T and 3 T) in human MRI scanners. When using scanner's gradient modulations, the technique is categorized as Wave-CAIPI [16] if a regular undersampling mask with CAIPIRINHA pattern shifts [71] is used, or CS-Wave [72] if a random mask is used. In the broader context, this work extends our recently proposed MRI framework for nonlinear gradient modulations, where the entire acquisition pipeline can be understood and formulated in k-space: signal encoding [30], field calibration [67], and now, joint data compression to reduce unnecessary memory expansion and enable compressed-sensing reconstructions. Together, these building blocks unify MRI spatial encoding by dynamic $B_0$ fields and static RF receive fields, turning the oversampled readout axis coupled with rapid $B_0$ field modulations into another acceleration dimension that could be mathematically-similar to parallel imaging. Generally, the proposed group-patch joint compression is universally-applicable to image acceleration based on rapid magnetic fields modulation techniques, regardless of the hardware used to generate the modulations (e.g., scanner gradients [16, 34, 41, 68, 69, 73–76], gradient inserts [55, 59], FRONSAC [28, 29], large matrix gradient coils [77], local $B_0$ coils [30, 32], and switched RF coils [78, 79]).

## 2 | Theory

### 2.1 | Overview: Hardware, Pulse Sequence, and Data Processing

As illustrated in Figure 1, FLASH line-by-line scans (2D and 3D) are accelerated by superimposing additional rapid $B_0$ field modulations onto the frequency-encoding gradient during signal readout (i.e., conventional frequency-encoding), using either a custom-built 8-channel local $B_0$ coil array placed outside the RF coil, or scanner's linear gradient for rapid modulations. Sinusoidal currents are applied to each channel of local $B_0$ coils or scanner gradients, imposing either local or global spin phase modulations, improving pixel disentanglement in image reconstruction using retrospectively undersampled k-space data. Prior to image reconstruction, the spatial encoding maps from dual auto-calibrating signal (ACS) scans [80] and undersampled imaging data are compressed using the proposed joint compression technique.

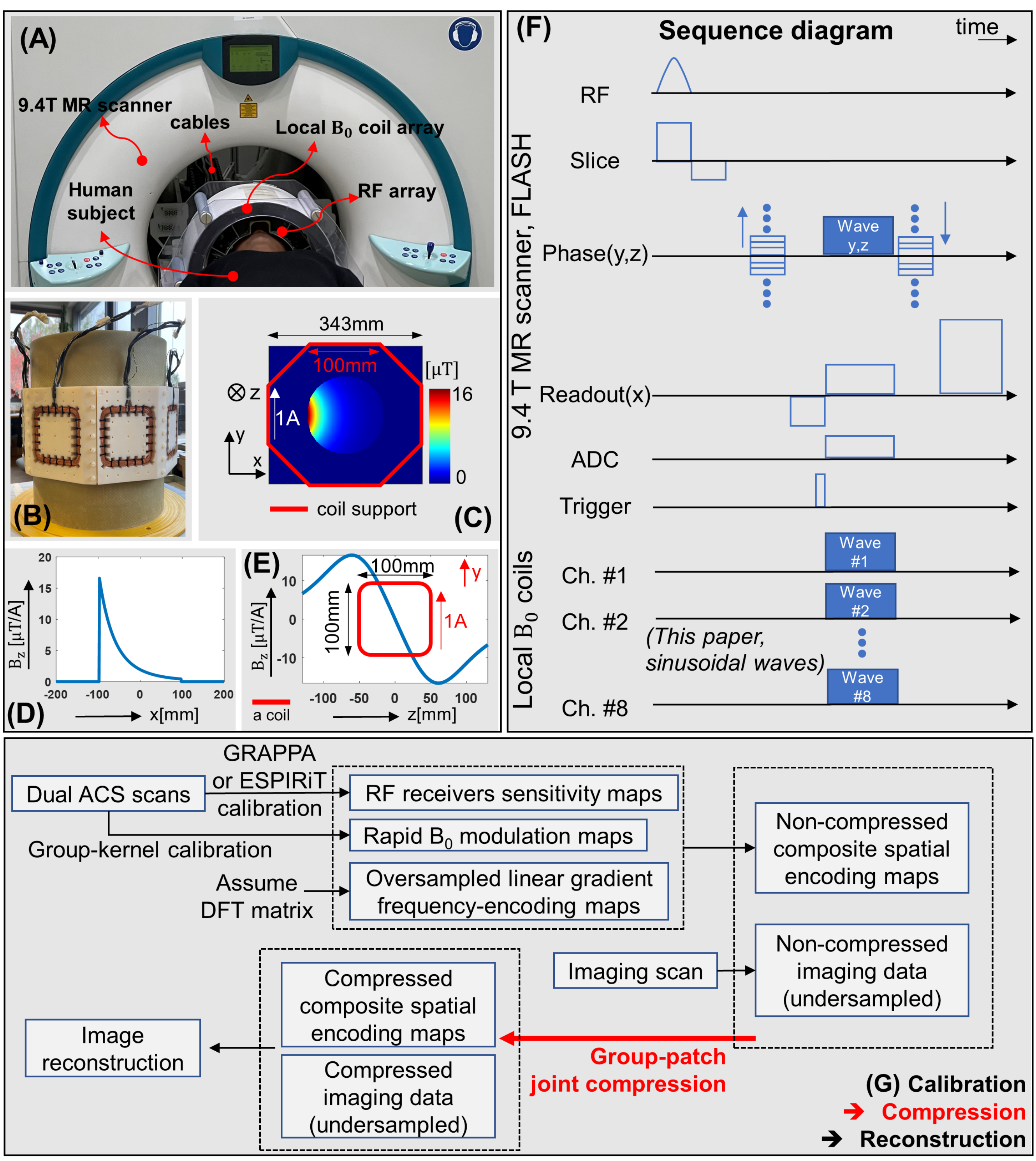

**FIGURE 1** | System illustration. (A) The hardware setup in the scanner. (B) The 8-channel local $B_0$ coil array. (C) Simulated magnetic field map given 1A DC current in a single local $B_0$ coil, which generates up to around 16 $\mu T$ within the effective field of view (FOV) defined by the inner boundary (around 97 mm to the FOV center) of an RF array. (D) The 1D plot along the central horizontal line in (C). (E) The variation of the in-plane maximal field strength generated by 1A in a local coil along the z direction, with the relative position to the coil winding indicated in red. (F) Pulse sequence diagram for accelerated MRI scans with rapid $B_0$ field modulations. In a standard FLASH sequence, during signal readout, either the local $B_0$ coil array or the scanner's phase-encoding gradients are switched on with current waveforms. Theoretically, the waveforms in the 8 local $B_0$ channels can be designed independently. Here, we apply sinusoidal currents for simplicity, to induce additional localized spin phase modulation on the objects to accelerate k-space acquisitions. (G) The data processing pipeline consists of calibration, compression and reconstruction steps. The group-patch joint compression proposed in this paper is highlighted in red.

## 2.2 | Signal Encoding in Image-Space and k-Space

In Cartesian MRI (in this paper, FLASH) accelerated by rapid $B_0$ field modulations and multiple RF receivers, the signal encoding model that maps image-space to k-space can be written as follows:

$$S_n(t) = \int c_n(r)\rho(r)exp\left\{-i\left[k(t)r + k_m(t,r)\right]\right\}dr, \tag{1}$$

$$k(t) = \gamma \int_0^t g(\tau)d\tau, \tag{2}$$

$$k_m(t,r) = \gamma \sum_{\lambda}\left[B_{\lambda}(r)\int_0^t I_{\lambda}(\tau)d\tau\right], \tag{3}$$

$S_n(t)$ is the time-domain signal obtained by the $n$th RF receiver coil at the time instant $t$. $k(t)$ is the k-space trajectory term (i.e., in rad/m) imposed by scanner's frequency and phase encoding, $k_m(t,r)$ is the phase evolution term imposed by the rapid $B_0$ gradient modulations (i.e., in rad). $\gamma$ is the gyromagnetic ratio (i.e., in rad/s/T), $r$ is the spin spatial location, $\rho(r)$ is the object spin density, $c_n(r)$ is the $n$th RF receiver sensitivity distribution, $g(\tau)$ is the shaped pulse of the linear gradients for the main trajectory (excluding rapid modulations), $B_{\lambda}(r)$ is the $B_0$ field distribution produced by unit current in the $\lambda$th local $B_0$ coil or scanner gradient element, $I_{\lambda}(\tau)$ is the current waveform in the $\lambda$th local coil or the scanner's linear gradient for rapid modulations only.

After discretization over space and stacking time samples and receiver channels:

$$S = E\rho, \tag{4}$$

$S \in \mathbb{C}^{N_t N_C \times 1}$ is the acquired k-space data vector with $N_t$ samples from each of the $N_C$ receivers. $E \in \mathbb{C}^{N_t N_C \times L}$ is the composite encoding matrix incorporating $B_0$ and RF encoding functions. $\rho \in \mathbb{C}^{L \times 1}$ is the image vector with $L$ pixels. The total sample number of each receiver is the product of the sample counts along the three dimensions, that is, $N_t = N_{t_x} N_{t_y} N_{t_z}$.

For a fixed acquisition time instant $t_0$, by the convolution theorem, the k-space signal in Equation (1) can also be seen as the convolution of the Fourier transforms of the three image-space maps, evaluated at a k-space location $k(t_0)$ or a sampling time instant $t_0$. This signal model maps between k-space data acquired without and with field modulation, via interpolation with time-varying $B_0$ and time-invariant RF kernels.

$$S_n(t_0) = \left\{ \underbrace{\mathcal{F}\left[c_n(r)\right]}_{RF\ receive\ sensitivity\ kernel} * \underbrace{\mathcal{F}[\rho(r)]}_{original\ k-space\ signal} * \underbrace{\mathcal{F}\left\{exp\left[-ik_m(t_0,r)\right]\right\}}_{rapid\ B_0\ modulation\ kernel} \right\}(t_0), \tag{5}$$

where asterisk $*$ is the convolution operation. Thus, the modulated sample $S_n(t_0)$ can be interpreted as a weighted sum of local neighborhoods of the original k-space data. The RF sensitivity induces a receiver-specific but time-invariant kernel, while the rapid $B_0$ modulation induces a receiver-invariant but time-varying kernel. For auto-calibration and interpolation reconstruction in parallel imaging, the multiple RF receive kernels with optionally omitted samples reduce to the well-known GRAPPA kernels.

The two-domain interpretations are illustrated in Figure 2. Here, two sets of k-space ACS regions are acquired without and with field modulations, which are used for $B_0$ field auto-calibration [67, 80]. Assuming that the ADC can sample faster than the Nyquist rate (determined by the FOV and gradient strength), the signal readout can be oversampled along time axis $t$ (specifically, $t_x$ in Figure 2). Without rapid $B_0$ modulations (i.e., $k_m(t,r) = 0$), this oversampling does not add new information; instead, if not averaged onto a non-oversampled grid, it simply increases the FOV along readout, introducing signal-free empty background. When switching on the $B_0$ modulation gradients during oversampling, the neighboring readout samples become informative because the modulation additionally mixes the phase-encoding dimensions. In Figure 2C–E, this is visualized by the kernel "cubes" in the standard ACS block, which depict the local support of the $B_0$ modulation kernels (reducing to an off-center point in the special case of purely linear gradient modulation). The kernel support—that is, the cube width or off-center displacement—reflects the extent of $B_0$-modulation-induced encoding along phase-encoding dimensions that speeds up acquisitions. Thus, rapid $B_0$ field modulations—regardless of whether linear (e.g., Wave-CAIPI) or nonlinear (e.g., FRONSAC, local $B_0$ coils modulations) gradients are used—modify an otherwise redundant oversampled readout to capture useful phase-encoding information.

## 2.3 | Multidimensional Compressibility and Periodic Redundancy

For reconstruction, although keeping the modulation scheme identical across phase-encoded steps allows efficient use of FFT for reconstructing phase-encoding dimensions [16, 32, 81], the remaining non-Fourier readout encoding model to be inverted scales with the product of the readout oversampling factor (e.g., ×8) and the number of RF receiver channels (e.g., ×32). This becomes even more computationally demanding in compressed-sensing with variable-density-Poisson-disc (VD-PD) [82] undersampling patterns, since the readout forward model becomes difficult to decompose into small groups of aliased phase-encoded pixels [74, 83].

However, the period of the oscillating $B_0$ fields typically ranges from 500 to 50 $\mu s$. As in Figure 2A,B, although the additional $B_0$ field modulation varies rapidly relative to the scanner's readout gradient (DC) used for conventional frequency-encoding, it still changes slowly relative to the scanner's oversampled ADC dwell time of a few $\mu s$. To exploit the resulting encoding redundancy across adjacent oversampling timepoints, the proposed group-patch joint compression technique jointly compresses small patches of neighboring readout samples across all RF receivers in groups. Leveraging the sinusoidal periodicity, we partition the readout axis into sub-intervals (corresponding to k-space different subregions) within one $B_0$ modulation period and assign each sub-interval a compression group index; the same set of indices repeats for every subsequent $B_0$ modulation period, so that samples at the same relative position within another modulation cycle can be compressed using the same group-specific compression matrix.

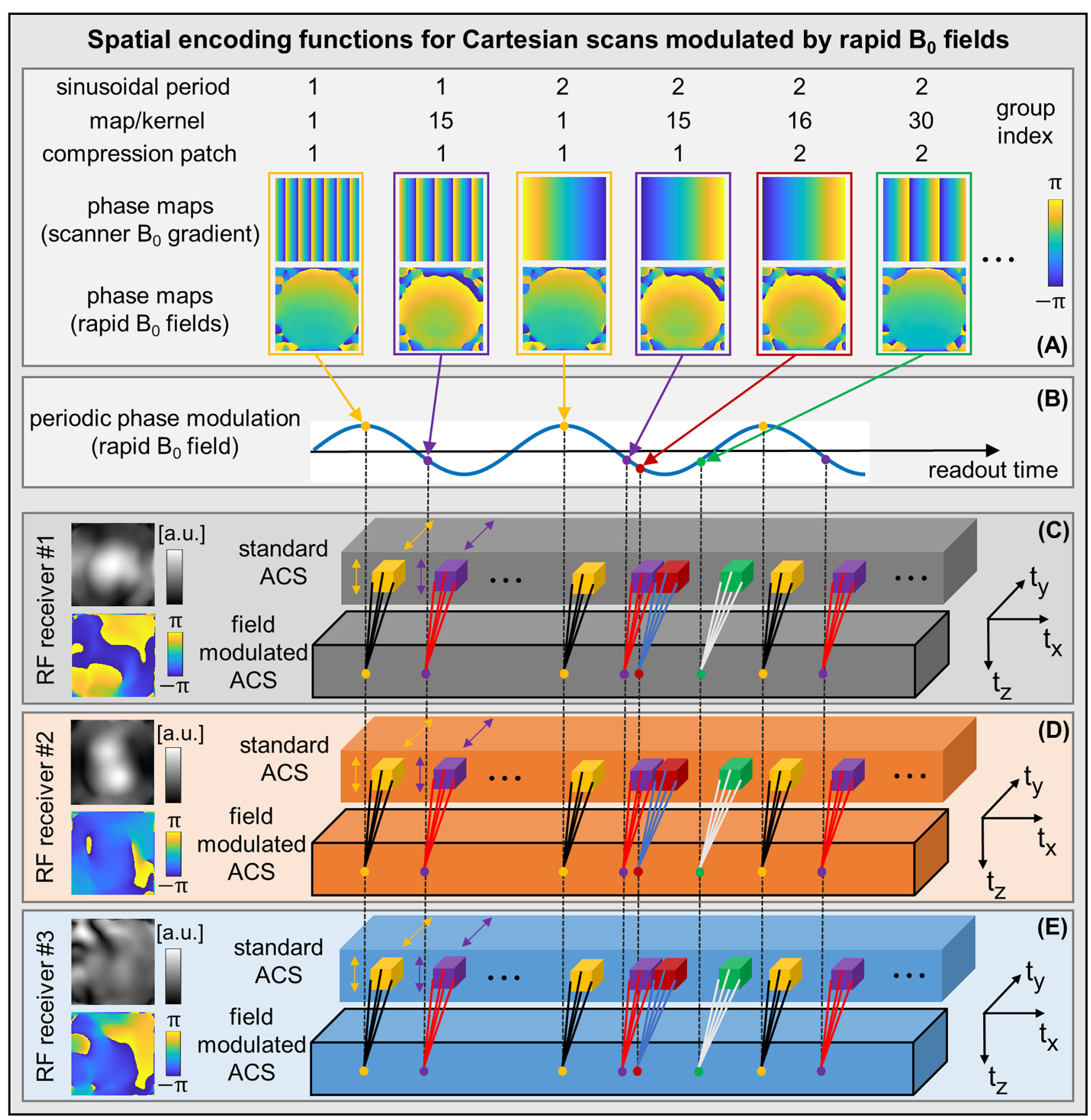

**FIGURE 2** | Signal encoding in image-space and k-space, accelerated by rapid $B_0$ modulation and static RF receivers sensitivities. (A) The spin phase maps induced by static (i.e., scanner frequency-encoding) and dynamic (i.e., additional modulation) $B_0$ fields, and the group indices for the $B_0$ kernels (auto-calibration purpose) and k-space subregions (compression purpose) along the readout time axis. (B) Sinusoidal modulation waveform illustrating the modulation phase of the oscillating $B_0$ fields, regardless of whether their spatial patterns are linear or nonlinear. (C-E) The k-space view of rapid $B_0$ field modulation, given three RF receiver channels. In this example, two fully-sampled ACS datasets are acquired to auto-calibrate the $B_0$ modulation kernels using GRE line-by-line scans. The standard ACS (no black outline) can also be used separately for parallel imaging calibration. The field-modulated ACS block (black outline) is acquired under additional rapid $B_0$ modulation; consequently, each sample can be expressed as the sum value of a weighted neighborhood (kernel) of the original k-space data in the non-modulated standard ACS block. The time-varying kernels are depicted by small cubes in different colors, indicating different modulation phases in a sinusoidal period. Different colors for ACS blocks denote different RF receive channels, with corresponding sensitivity maps shown. Auto-calibration of the phase evolution maps imposed by rapid $B_0$ modulation is performed by solving these k-space interpolation relationships within each kernel group. Note that, both the auto-calibration and the joint compression exploit the periodicity of rapid $B_0$ modulations, but with different ways of grouping. Generally, the time-varying $B_0$ kernels can be partitioned into time-invariant subsets—either time-invariant kernels or time-invariant encoding subregions—by grouping samples that share the same sinusoidal modulation phase across the entire k-space.

Separately, $B_0$ modulation kernels can also be assigned kernel group indices for the auto-calibration purpose [67, 80]. These kernel indices are defined per readout time instant and repeat across sinusoidal cycles, and therefore differ from the compression grouping that aggregates multiple neighboring readout samples into a patch. Nevertheless, for a given modulation phase instant or segment, the $B_0$ encoding functions are shift-invariant across modulation cycles.

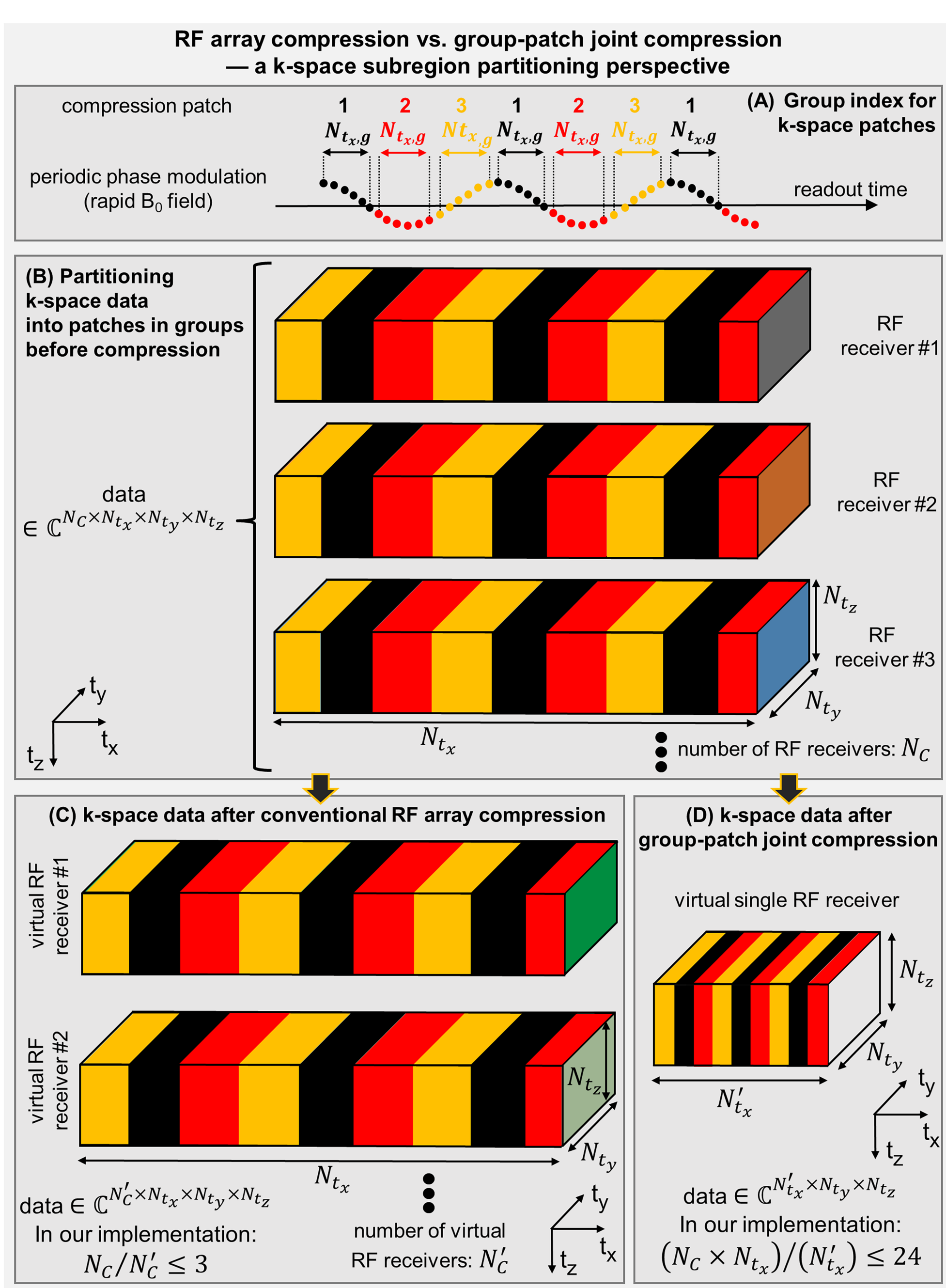

**FIGURE 3** | Comparison between conventional RF array compression and the proposed group-patch joint compression, from the perspective of k-space subregion partitioning. (A) Group indices defining k-space compression patches/subregions. The readout time axis is divided into segments spanning, for example, $N_{t_x,g}$ timepoints/samples. Because of the periodicity of rapid $B_0$ modulation, each sinusoidal period can be partitioned into a small number of segments governed by distinct $B_0$ encoding functions, which repeat across consecutive sinusoidal periods. (B) Visualization of the k-space data volume prior to compression, showing readout oversampling in different k-space subregions (black, red, yellow) and multiple RF receiver channels (gray, orange, blue). (C) Data volume after conventional RF array compression. The receiver dimension is reduced while all other k-space dimensions remain unchanged. In our implementations, to avoid loss in encoding efficiency, the compression factor (RF-only) may need to stay below 3×. (D) Data volume after the proposed joint compression. The RF receive channels and the readout subregions are compressed jointly, yielding dimensionality reduction along both receiver and readout dimensions. In our implementation, up to 24× compression factor (joint compression) could be possible without noticeable loss in encoding efficiency.

## 2.4 | From RF Array to Joint $B_0$-RF Compression

To explicitly visualize how the compressed spatial encoding maps evolve with increasing compression factors, we compute group-specific compression matrices by feeding composite spatial encoding maps into a conventional sensitivity-based RF array compression method [38]. Now, these maps incorporate both $B_0$ (frequency-encoding, rapid modulation) and RF (receive) fields that jointly encode each subregion group. By treating the joint $B_0$ and RF spatial encoding maps as an expanded set of virtual receiver maps, this implementation also inherits the advantage of this array compression algorithm based on explicit receiver maps and yield the optimized compression matrices computed from the encoding maps within the region-of-interest (ROI). In contrast, as in Figure 3, the RF-only compression reduces only the receiver-channel dimension, whereas our joint $B_0$-RF compression compresses both the receiver and segmented-readout dimensions using a group-specific operation. Theoretically, joint compression may also be performed using k-space data alone by adapting alternative receiver-compression algorithms [39, 40, 42] to the same subregion-wise formulation.

While a full derivation of the original RF array compression method can be found in reference [38], we summarize our compression steps for each k-space subregion group $g$ as below. These steps are repeated for different compression groups, that together span full modulation cycles across the entire k-space.

1. Form an ROI-aggregated square matrix $P_g$ from the composite spatial encoding maps.
2. Compute a compression matrix $A_g$ from $U_g^H$ and the receivers noise-decorrelation matrix $T$. $U_g^H$ is obtained via singular-value decomposition (SVD) or eigenvalue decomposition of the matrix $P_g$.
3. Apply $A_g$ to compress the corresponding k-space patches and composite spatial encoding maps, which are then used for image reconstruction.

First, we define a composite encoding vector at a pixel $r \in ROI$, cropped from $E$ in Equation (4):

$$E_g[r] = E_{LG,g}[r] \odot E_{NG,g}[r] \odot E_{RF}[r] \in \mathbb{C}^{N_s \times 1}, \tag{6}$$

where $N_s = N_C N_{t_x,g}$, $N_{t_x,g}$ is the number of oversampled readout timepoints within one $B_0$ modulation cycle in the group $g$, $N_C$ is the total receiver channels. $E_g[r]$ is constructed by elementwise multiplication $\odot$ of all spatial encoding vectors at this pixel, induced by the scanner readout gradient ($E_{LG,g}[r] \in \mathbb{C}^{N_s \times 1}$), local $B_0$ coils ($E_{NG,g}[r] \in \mathbb{C}^{N_s \times 1}$) and RF sensitivity maps ($E_{RF}[r] \in \mathbb{C}^{N_s \times 1}$) (optionally noise-decorrelated, repeated across k-space locations). For RF-only compression, the Equation (6) reduces to $E_g[r] = E_{RF}[r]$.

Consequently, we form the square matrix $P_g \in \mathbb{C}^{N_s \times N_s}$, by taking the outer product between the composite encoding column vector and its pseudo-inverse row vector, and summing over $L$ spatial pixels within the ROI (e.g., image mask).

$$P_g = \sum_{r \in ROI} E_g[r] E_g^{\dagger}[r], \tag{7}$$

$$= U_g F_g U_g^H, \tag{8}$$

where $(\cdot)^{\dagger}$ is the pseudo-inverse operation that leads to the optimized compression due to orthonormalization of the encoding vectors [38]. $(\cdot)^H$ denotes conjugate transpose. The unitary matrix $U_g^H$ contains the eigenvectors of $P_g$, which can be obtained via SVD/eigenvalue decomposition in practice.

By the definition of pseudo-inverse $E_g^{\dagger}[r] = \left(E_g^H[r] E_g[r]\right)^{-1} E_g^H[r]$, since $E_g[r]$ is a column vector, $\left(E_g^H[r] E_g[r]\right)^{-1}$ is a scalar. Thus, we can compute $P_g$ efficiently by rewriting it as direct matrix multiplication operations, without the time-consuming explicit per-pixel vector inversion and the subsequent spatial summation:

$$P_g = \sum_{r \in ROI} \frac{E_g[r]}{E_g^H[r] E_g[r]} E_g^H[r], \tag{9}$$

$$= \left(E_g N\right) E_g^H, \tag{10}$$

$$\text{with } N = diag\left[1 / \left(\sum_{N_s} E_g \odot E_g^*\right)\right], \tag{11}$$

where $E_g \in \mathbb{C}^{N_s \times L}$ is the composite spatial encoding maps for group $g$, that span $N_s$ k-space positions and $L$ pixels, $(\cdot)^*$ is the complex conjugate operation, $N \in \mathbb{C}^{L \times L}$ is a diagonal matrix with a scaling vector $1 / \left(\sum_{N_s} E_g \odot E_g^*\right) \in \mathbb{C}^{1 \times L}$ at its diagonal for pixel-dependent normalization. This scaling vector is obtained by taking elementwise square operation of $E_g$, summing over along k-space and receiver dimensions as $\sum_{N_s} E_g \odot E_g^*$, and taking the inverse.

Second, a compression matrix $A_g \in \mathbb{C}^{N_s' \times N_s}$ can be obtained:

$$A_g = C_g U_g^H T, \tag{12}$$

where $T$ is the receiver noise-whitening transform (e.g., obtained from Cholesky factorization of the noise covariance matrix and inversion of the decomposed $Q$, as $\Psi = QQ^H \rightarrow T = Q^{-1}$), $U_g^H$ is the unitary matrix obtained via e.g., SVD or eigenvalue decomposition of the matrix $P_g$ as in Equation (8), and $C_g$ is a selection matrix that retains the first $N_s'$ rows of $U_g^H$. The singular-value (or eigen-value) spectrum from $F_g$ in Equation (8) can quantify the compressibility of the spatial encoding functions, indicating a reasonable value of $N_s'$ to preserve the essential information.

Third, the compression matrix $A_g$ is used to compress the k-space data patches $S_g \in \mathbb{C}^{N_s \times G_{t_x,g} N_{t_y} N_{t_z}}$ across all sinusoidal modulation cycles along readout and phase-encoded steps, and the corresponding composite spatial encoding maps $E_g$ belonging to this compression group $g$. $G_{t_x,g}$ is the repetition time of the group $g$ along readout time axis. $N_{t_y}$ and $N_{t_z}$ are the phase-encoded steps along two dimensions, respectively. This compression matrix can be efficiently reused across different modulation cycles, since these k-space subregions are encoded by the identical spatial modulations caused by rapid $B_0$ and RF receivers fields, and only differs in the k-space locations.

$$S_g' = A_g S_g, \tag{13}$$

$$E_g' = A_g E_g, \tag{14}$$

The resulted compressed k-space data $S_g' \in \mathbb{C}^{N_s' \times G_{t_x,g} N_{t_y} N_{t_z}}$ and the virtual spatial encoding maps $E_g' \in \mathbb{C}^{N_s' \times L}$ in different groups are combined into a reduced forward model as below, used for least-squares [16, 30, 32, 81] or compressed-sensing [34] image reconstruction. The RF array compression and the group-patch joint compression are visually illustrated in Figures 4 and 5.

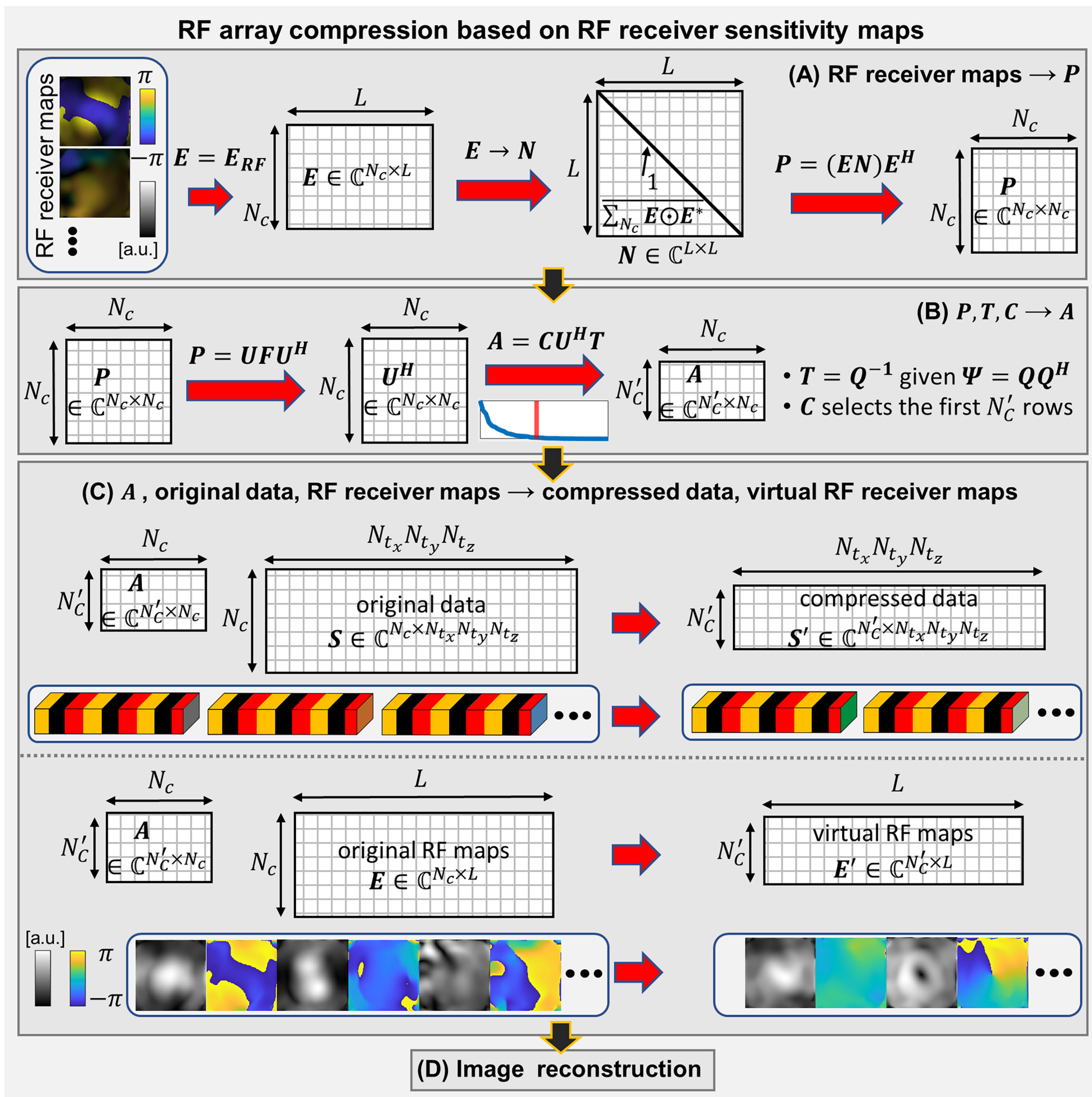

**FIGURE 4** | Visualization of the conventional RF array compression, using similar notations for the joint compression but without group index $g$. (A) Construction of the square matrix $P$ from RF receiver maps $E_{RF}$, with $N_C$ receivers and $L$ spatial pixels. Note that the RF sensitivity maps are displayed with overlaid amplitude (grayscale) and phase (parula) to save figure space. Amplitude variations are indicated by the non-uniform darkness within the maps. (B) Calculation of the compression matrix $A$ from $P$, the noise decorrelation matrix $T$, and the selection matrix $C$. Namely, an SVD on $P$ with subspace thresholding by $C$ is utilized to yield the compression matrix for RF sensitivity compression alone. $N_C$ receivers will be compressed to $N'_C$ virtual receivers. (C) Application of $A$ to compress the k-space data and sensitivity maps along receiver dimension only. Colored k-space subregions are shown to emphasize that no dimensionality reduction is performed along the readout axis. (D) The compressed k-space data and the compressed RF receiver maps are loaded into the image reconstruction pipeline to reconstruct the final image with less computational time.

$$S' = E'\rho, \tag{15}$$

where $S' \in \mathbb{C}^{N'_{t_x}N_{t_y}N_{t_z}\times 1}$ and $E' \in \mathbb{C}^{N'_{t_x}N_{t_y}N_{t_z}\times L}$. $N'_{t_x} = N'_s G_{t_x,g}$ is the compressed virtual dimension of the joint readout-RF receivers space.

## 3 | Methods

### 3.1 | Hardware Setup and Coil Configuration

MRI experiments were performed on a 9.4 T whole-body human scanner equipped with SC72 gradients (max. 200 T/m/s, max. 70 mT/m), and a 3 T whole-body Prisma$^{fit}$ human scanner equipped with XR gradients (max. 200 T/m/s, max. 80 mT/m) (Siemens Healthineers, Erlangen, Germany). At 9.4 T, for non-linear gradient modulation scans, an 8-channel local $B_0$ coil array [30] was fixed on the patient table, surrounding the image region of interest. A 16-transceive/32-receive RF coil array, modified from the reference [84] and using circularly polarized (CP) mode for RF transmission, was placed inside the local $B_0$ coil array's hollow cylindrical support. For Wave-CAIPI/CS-Wave using scanner gradient modulations, the local $B_0$ coil array is switched off. At 3 T, 34 or 44 elements of a 64-channel RF receive array were used.

### 3.2 | Pulse Sequence and Data Acquisition

Line-by-line 2D/3D FLASH-based sequences (modified IDEA sequence for simultaneously driving local $B_0$ coils [30], Pulseq sequence for Wave-CAIPI/CS-Wave) were used for ex vivo and in vivo scans, with 8× readout oversampling and 3 $\mu s$ scanner ADC dwell time. Several healthy adults volunteers were scanned, in agreement with the institution's ethics policy. During the signal readout (i.e., scanner's frequency-encoding), the local $B_0$ coil array or scanner's phase-encoding gradients were driven by independent sinusoidal modulations of each channel. Identical rapid $B_0$ modulations were used across TRs for simplicity. The optimized modulation schemes in previous studies [30, 67] were used to maximize acceleration capability of local $B_0$ coils. The nonlinear gradient strengths and the least-squares encoding efficiency maps (e.g., G-map, power function maps) of the in vivo scans are presented in Supporting Information and reference [30]. In 2D coronal scans, the local $B_0$ coils produced a nearly-linear z-gradient as bunched phase encoding (BPE). In 3D scans, they produced a quadratic function in transverse plane coupled with a nearly-linear gradient along $z$, as a form of nonlinear gradient modulations. Fully-sampled low-resolution FLASH with and without rapid $B_0$ modulations were acquired as dual ACS, to calibrate $B_0$ field modulations via $B_0$ kernel model [80]. Fully-sampled imaging datasets were acquired and retrospectively undersampled. Note that, this helps evaluating reconstructions using various k-space undersampling masks conveniently based on the same dataset, but can increase susceptibility to motions during lengthy scans.

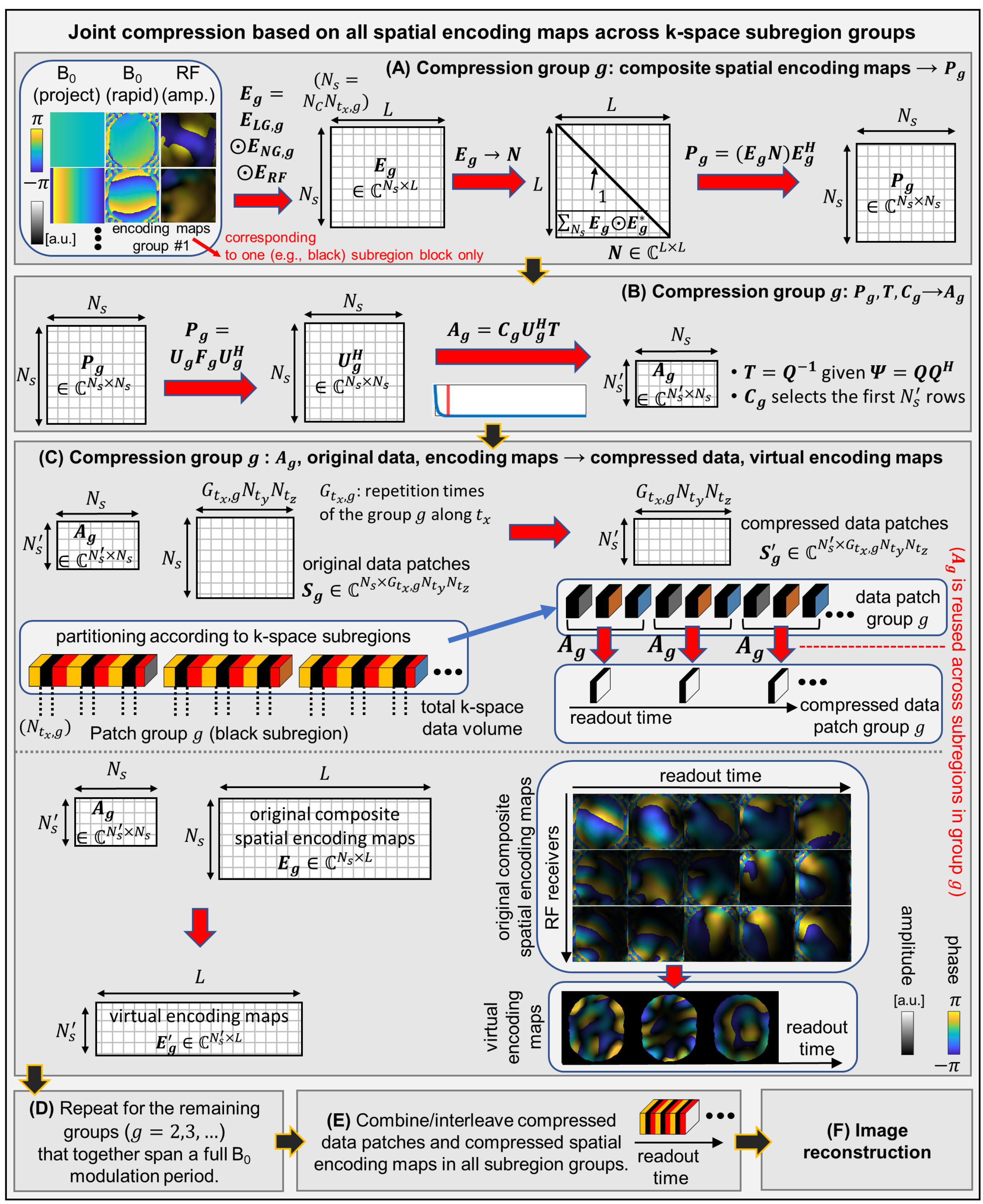

**FIGURE 5** | Visualization of the proposed group-patch joint compression. (A) Construction of the square matrix $P_g$ from all spatial encoding maps associated with the subregion group $g$, with $N_s$ readout samples from all receivers and $L$ spatial pixels. (B) Calculation of the group-specific compression matrix $A_g$ from $P_g$ and the noise decorrelation matrix $T$. An SVD on $P_g$ with subspace thresholding by $C_g$ is applied to compress the composite signal encoding space, from $N_s$ samples to $N_s'$ samples. (C) Application of $A_g$ to jointly compress the k-space data and composite spatial encoding maps along both receiver and readout dimensions. For a given sinusoidal modulation cycle, k-space patches belonging to the group $g$ (here, the black subregion) from different receiver channels (gray/orange/blue) are compressed together. Patches from the same group occurring in other modulation cycles and across phase-encoded dimensions are compressed using the same compression matrix, avoiding the need to compute a separate compression matrix at a different k-space location. (D) This procedure is repeated for all subregion groups. (E, F) The resulting compressed patches and compressed composite encoding maps are then interleaved/combined to form a reduced signal encoding model for faster image reconstruction.

Essential sequence protocols are summarized below; additional details are in the Supporting Information. In Figures 7 and S1, the 2D multi-slice ex vivo scan at 9.4 T has a matrix-size of $180 \times 180$ ($1.1 \times 1.1\,\text{mm}^2$). In Figure S2, the 3D ex vivo scan at 9.4 T has matrix-size of $180 \times 180 \times 80$ ($1.1 \times 1.1 \times 1.0\,\text{mm}^3$). In Figures S3 and S4, the 2D multi-slice in vivo scan at 9.4 T has a matrix-size $280 \times 280$ ($0.89 \times 0.89\,\text{mm}^2$). In Figures 8 and S4, the in vivo 3D scan at 9.4 T has a matrix-size $230 \times 230 \times 112$ ($1.0 \times 1.0 \times 1.0\,\text{mm}^3$). In Figure 9, the in vivo 3D scan at 3 T has a matrix-size $230 \times 230 \times 144$ ($1.0 \times 1.0 \times 1.0\,\text{mm}^3$). The TE/TR is 4–6.07/30 ms at 9.4 T, 20/35 ms at 3 T. The readout bandwidths are 150–230 Hz/Pixel. The ADC dwell time is 3 $\mu s$. The sinusoidal currents in local $B_0$ coils are $36\text{A}_{\text{pk}}$–$40\text{A}_{\text{pk}}$ per channel, 7.41 kHz. The Wave-CAIPI/CS-Wave has 24 wave-cycles with wave-gradients in y-3.4-z-5.3 mT/m and 4.17 kHz at 9.4 T, and 19 wave-cycles with y-6.4-z-7.7 mT/m and 3.33 kHz at 3 T. The 2D slice-thickness is 3 mm. The 3D slice oversampling is 26.3%–33.3%.

## 3.3 | Two Data Compression Approaches

For comparison, the conventional RF array compression [38] was performed as the baseline, based on the RF sensitivity maps estimated from standard ACS data (GRE) via ESPIRiT [24]. In the proposed joint compression approach, the oversampled readout axis was subdivided into subregions, each spanning 5–16 neighboring data points with all receiver channels. For each subregion, the composite encoding matrix is obtained from the synthesized DFT matrix, the rapid $B_0$ phase evolution maps auto-calibrated via k-space kernels [67], and the RF sensitivity maps used for RF array compression. Typically, one $B_0$ sinusoidal modulation period spans 45 readout timepoints using local $B_0$ coils, and 80–100 timepoints using the scanner's modulation gradients as in Wave-CAIPI/CS-Wave. Since identical rapid $B_0$ modulation was used across phase-encoded steps, k-space data acquired in different TRs can be compressed by the same group-specific compression matrices.

## 3.4 | Two Image Reconstruction Algorithms

As illustrated in Figure 6, we implemented two reconstruction algorithms offline in MATLAB, on a CPU node (2× AMD EPYC 7452, 2.35GHz, 64 cores, 2 hyper-threads/core, with 1 TB RAM option). Both algorithms can reconstruct an image $\hat{\rho} \in \mathbb{C}^{N_x \times N_y \times N_z}$ from the compressed samples $S' \in \mathbb{C}^{N'_{t_x} \times N_{t_y} \times N_{t_z}}$ (or the original samples $S \in \mathbb{C}^{N_C \times N_{t_x} \times N_{t_y} \times N_{t_z}}$ for the uncompressed baseline). Because of the identical rapid $B_0$ modulations across phase-encoding steps, the phase-encoding dimension(s) can be reconstructed efficiently using FFT in both pipelines.

In the reconstruction algorithm #1 (hybrid-space reconstruction [16, 30, 32, 81]), the phase-encoding undersampling is restricted to regular/CAIPI-type patterns that yield structured aliasing (i.e., an $R$-fold superposition of discrete pixels in the image-space of phase-encoding dimensions). Consequently, the global forward model in Equation (15) (mapping $\mathbb{C}^{N_x \times N_y \times N_z} \rightarrow \mathbb{C}^{N'_{t_x} \times N_{t_y} \times N_{t_z}}$) decouples into a set of small readout forward models (i.e., time axis only includes readout) that can be solved as independent unfolding subproblems. Each readout forward model maps all readout pixels and one aliased set of phase-encoding pixels as unknowns, to a single k-space readout line acquired from each RF receiver within one TR (mapping $\mathbb{C}^{N_x R} \rightarrow \mathbb{C}^{N'_{t_x}}$). Our implementation steps are summarized as follows to solve:

$$\hat{\rho} = \underset{\rho}{argmin} \left\| E' \rho - S' \right\|^2, \tag{16}$$

1. Apply inverse FFTs along phase-encoding dimensions to the k-space data.
2. For each aliased set of phase-encoding pixels, solve the corresponding readout forward model. These subproblems can be solved in parallel (e.g., MATLAB "parfor" with multi-threading). The LSQR [85] was used as the numerical solver.
3. Assemble the solutions from all aliased sets to form the full image.

In the second reconstruction algorithm #2, arbitrary phase-encoding undersampling patterns (PD-VD masks generated by BART [82]) can be used, which consequently enables advanced iterative reconstructions (e.g., compressed-sensing [19]) for higher acceleration factors. In this setting, aliasing in the phase-encoding dimension(s) can be incoherent and therefore may not yield separable aliased pixel sets after inverse FFT. As a result, the reconstruction must solve the global forward model in Equation (15) (mapping $\mathbb{C}^{N_x \times N_y \times N_z} \rightarrow \mathbb{C}^{N'_{t_x} \times N_{t_y} \times N_{t_z}}$), with all phase-encoded pixels estimated simultaneously. This increases the problem size and makes the reconstruction algorithm #2 computationally slower than the reconstruction algorithm #1. For PD-VD masks, the nominal undersampling factor is defined as the ratio between the number of sampled k-space locations and the total number of grid points on the full square Cartesian grid (without circular cropping). For the reconstruction algorithm #2, FISTA [86] was used as the numerical solver that

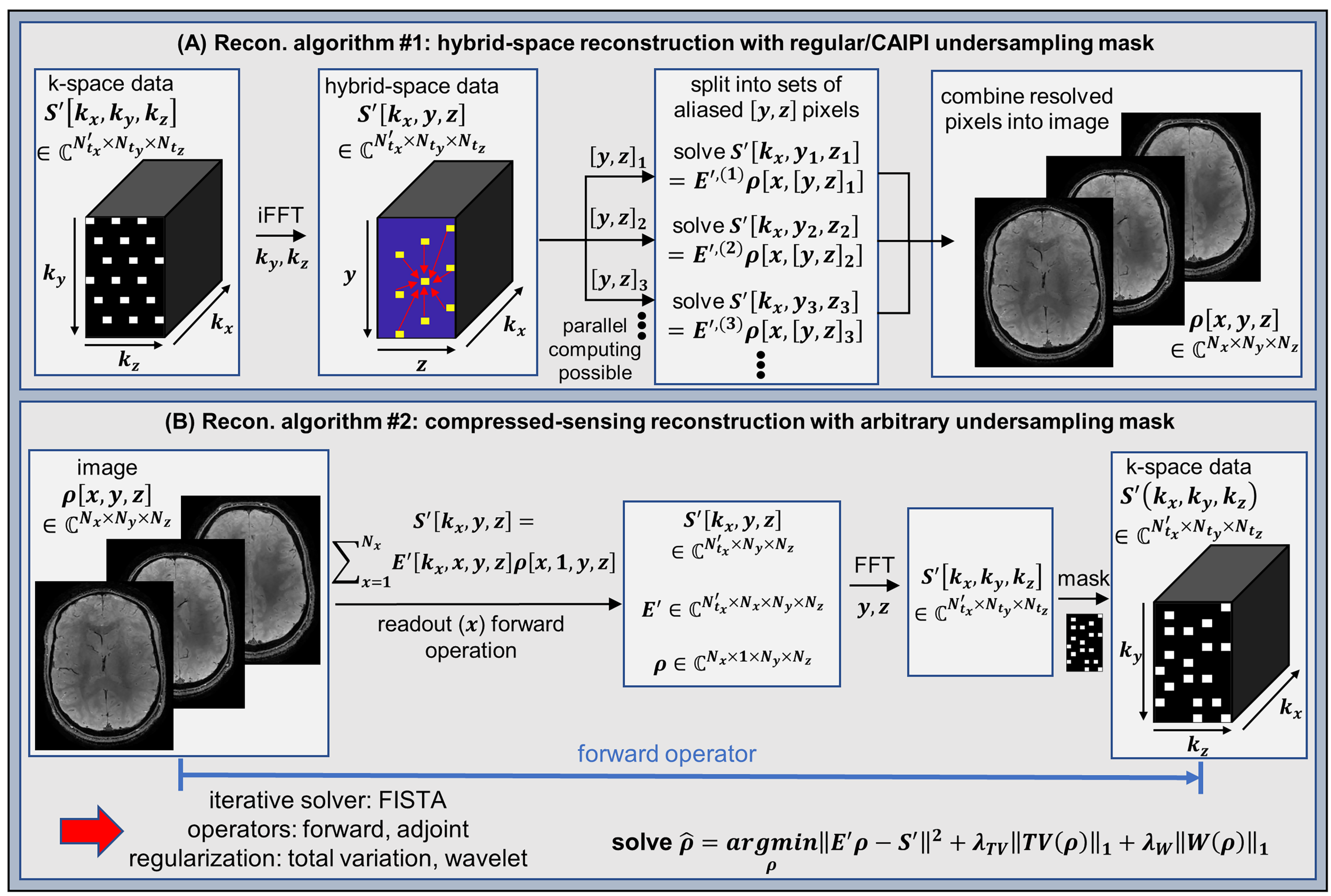

**FIGURE 6** | Reconstruction strategies. (A) Reconstruction algorithm #1. For regular/CAIPI undersampling, the global forward model decouples into independent small subproblems that can be solved efficiently in hybrid-space. $[y,z]_1, [y,z]_2, [y,z]_3$ are the superimposed phase-encoded pixels to be separated in the reconstructed image, which will be estimated separately in each set of the subproblems. (B) Reconstruction algorithm #2. For arbitrary undersampling (including regular/CAIPI as a special case), the global forward model is solved directly using FISTA. The corresponding forward operator is illustrated. Because this formulation generally supports arbitrary sampling masks, we implement our compressed-sensing reconstruction with total variation and wavelet regularization based on this scheme.

solves the objective:

$$\hat{\rho} = \underset{\rho}{argmin}\left\| E'\rho - S' \right\|^2 + \lambda_{TV}\|TV(\rho)\|_1 + \lambda_W\|W(\rho)\|_1, \tag{17}$$

where $TV(\cdot)$ denotes the total variation operation, $W(\cdot)$ denotes a wavelet transform. As a special case, algorithm #2 can also support regular/CAIPI undersampling pattern as a special case, where setting $\lambda_{TV} = \lambda_W = 0$ can reduce Equation (17) to least-squares reconstruction. The wavelet regularization was implemented by the soft-thresholding, and the total variation regularization was implemented by smooth approximations in practice. Theoretically, the auto-calibrated $B_0$ modulation kernels/maps should be robust to system imperfections (e.g., gradient timing delay, gradient nonlinearity, and eddy currents), thereby minimizing estimation errors in encoding matrix and reconstruction artifacts. However, in strongly-modulated, highly-undersampled scans, regularization is usually helpful for attenuating residue artifacts arising from subtle model mismatch, noise and undersampling.

To evaluate the reconstruction quality, the reconstructed image was subtracted by the reference image without compression and undersampling, and then divided by the reference image to yield a difference image in percentage. Further, the normalized-root-mean-square-error (NRMSE) was calculated comparing the reconstructed image and the reference image, normalized by the range of the reference image (i.e., maximum − minimum). Lastly, structural similarity index (SSIM) was computed comparing the reconstructed image and the reference one using a MATLAB built-in function.

## 4 | Results

### 4.1 | Visualization of Compressed Spatial Encoding Functions

Using ex vivo brain scans (eliminating concerns about motions), we first evaluate the compressibility of static RF receive and dynamic $B_0$ encoding functions, based on singular-value distributions of square matrix $P_g$ that spans the subspace of the composite encoding functions to be compressed, as Equations (7) and (8). First, the subspace of conventional RF array compression using only RF sensitivity maps is shown in Figure 7A. Second, Figure 7B shows the subspace of joint compression based on all spatial encoding maps across a k-space readout subregion within one $B_0$ modulation period, as assigned to a compression group.

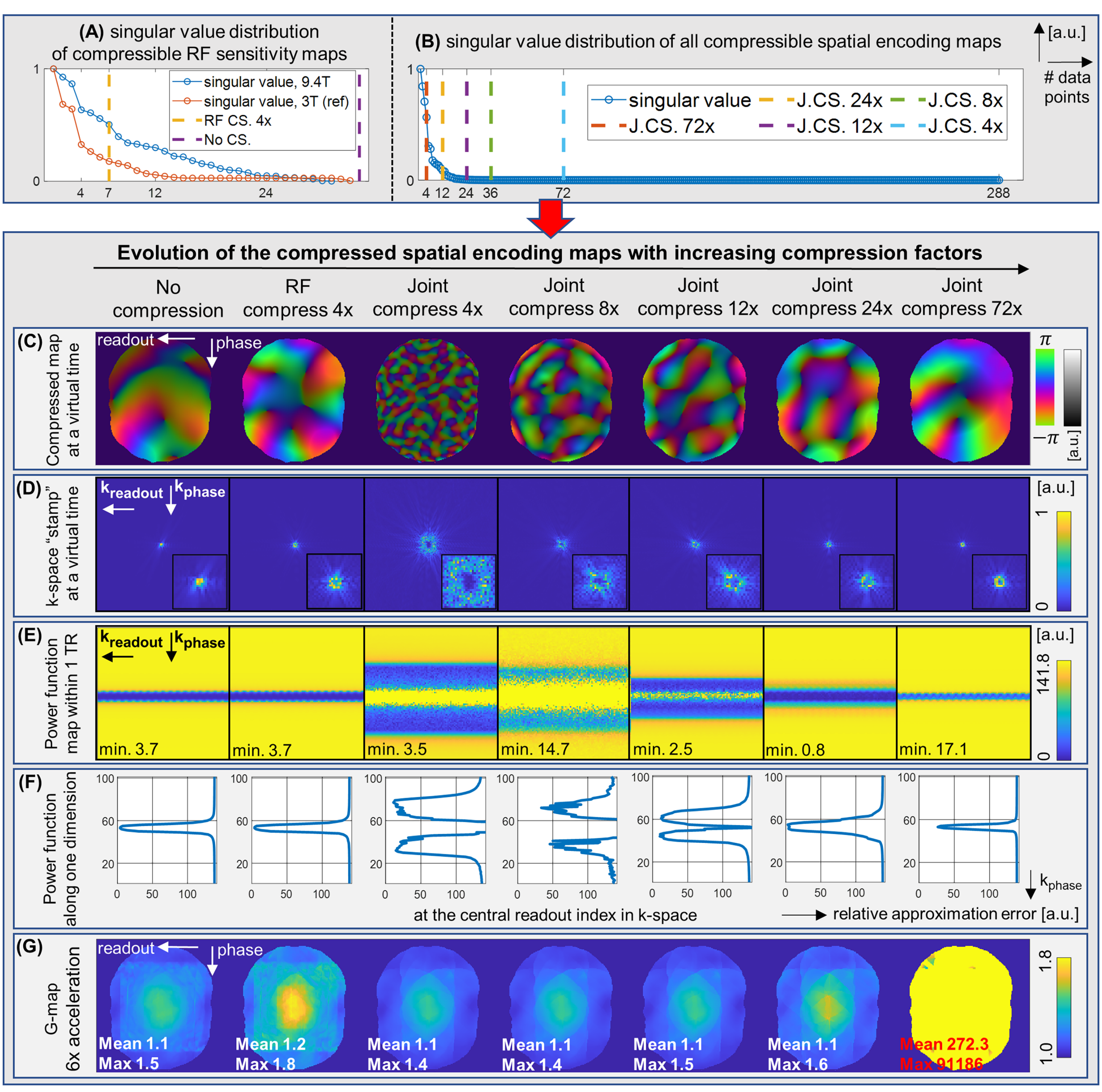

**FIGURE 7** | Visualizing the data compressibility of the RF-only and the joint compression approaches. (A) The singular value distribution of the SVD for the square matrix $P$ formulated based on RF sensitivity maps at 9.4 T and 3 T. Note, the plot excludes the first singular value which is way larger than others. (B) The singular value distribution of the SVD for the matrix $P_g$ constructed based on a composite encoding matrix including all spatial encoding maps across a readout duration. (C) The non-compressed or compressed maps at a (virtual) time, around the k-space center. (D) The k-space "stamps" or kernels as the Fourier transform of the non-compressed or compressed spatial encoding maps around the k-space center. (E) The power function maps as k-space approximation error for arbitrary spatial encoding functions. (F) The 1D plots of the power function maps at the central readout pixel in k-space. (G) The G-maps showing image-space noise amplifications given the non-compressed or compressed readout encoding matrix, with 6× phase undersampling.

For nearly-lossless RF array compression, these receive coils [84] at 9.4 T can be compressed from 32 to ~24 virtual coils, and the ones at 3 T can be compressed from 34 to ~15 virtual coils. By contrast, the more complex standing waves at 9.4 T improve the acceleration capabilities, but meanwhile, reduce the compressibility.

At 9.4 T, when extending the same compression algorithm to all spatial encoding maps, the total data can be compressed nearly-lossless by about 12×–24× folds. This surprisingly huge compressibility originates from the encoding redundancy introduced by oversampling the readout (8×) for capturing k-space signals modulated by $B_0$ fields that vary slowly relative to ADC dwell time.

Figure 7C–G illustrates the progression of compressed spatial encoding functions (image-space maps or k-space kernels) with increasing compression factors from left to right columns, based on a 2D ex vivo scan on the transverse plane. Additionally, it shows the corresponding k-space $B_0$ modulation kernel [49] at a (compressed) "time instant," the 2D power function maps representing k-space approximation errors within one phase-encoded step [30], the 1D distribution of the power function maps, and the G-map representing image-space noise amplification [10]. Here, the power function maps are used to evaluate the cumulative k-space sampling efficiency of arbitrary spatial encoding functions within a single TR, namely, the cumulative encoding effects given the dynamic kernels across one readout line. Such efficiency maps are derived from non-compressed or compressed signal encoding maps spanning one TR (e.g., cropped from $E$ or $E'$), utilizing Equations (14–16) in our recent nonlinear gradient sampling theory based on reproducing kernel Hilbert space (RKHS) [30]. From this view of RKHS (a Hilbert space of encoding functions), the arbitrary spatial encoding functions under investigation can be mathematically interpolated onto a fully-sampled Fourier function baseline. The power function characterizes the time-domain approximation error of this interpolation, obtained from the subtraction between the Gram matrix of the ideal fully-sampled DFT baseline and the cross-Gram matrix between the investigated encoding functions and this DFT baseline. Consequently, lower power function values reflect a smaller approximation error during interpolation (yielding higher encoding efficiency), whereas higher values indicate the opposite.

The 4× RF-only compression minimally perturbs the encoding functions, but raises the maximum G-factor to 1.8, indicating compromised sampling efficiency. In contrast, the joint compression can abruptly reshape the virtual spatial encoding maps or their equivalent k-space kernels ("stamps"), far more than RF-only compression, yet preserve overall encoding efficiency across phase-encoded steps (G-map) up to about 24× compression. As the joint compression factor increases, the effective k-space coverage (power function) per phase-encoded step evolves from a single trough to two troughs (reflecting redistributed k-space support), then gradually returns toward a single trough at higher factors, as the SVD numerically selects the most informative encoding modes. Note, some compressed spatial encoding functions cannot be physically generated by our current hardware directly, but are merely numerical consequences.

Note, for 8× joint compression in Figure 7E,F, larger k-space approximation errors with wider support per TR do not necessarily lead to degraded overall sampling efficiency considering all phase-encoded steps.

In Supporting Information, Table S1 summarizes the transitioning characteristics for different compression types and increasing compression factors in Figure 7C–G. In Figures S1 and S2 and Table S2, this ex vivo brain phantom was also used to investigate the speed and quality for each of the compression-reconstruction steps, given different compression factors, undersampling masks, and reconstruction algorithms.

### 4.2 | RF-Only Versus $B_0$-RF Compression for In Vivo 3D Scans at 9.4 T

In Figure 8, with reasonable compression factors at 9.4 T, reconstructions of 3D in vivo FLASH with rapid modulations of local $B_0$ coils (least-squares $L_2$ recon with regular/CAIPI patterns, CS recon with random patterns) or scanner's gradients (least-squares $L_2$ Wave-CAIPI/CS-Wave) are shown, comparing the RF-only compression and the proposed group-patch joint compression. A similar study but for 2D FLASH with rapid $B_0$ modulations is documented in Supporting Information (Figure S3), showing 20× joint compression that substantially accelerates 2D compressed-sensing reconstruction.

As in Figure 8A–H, for nonlinear-gradients-modulated scans, least-squares reconstruction (using reconstruction algorithm #1) with 4 × 3 undersampling and compressed-sensing reconstruction (using reconstruction algorithm #2) with 14.6× PD-VD undersampling are shown. For $L_2$ reconstructions, 18× joint compression took noticeably more compression time than 4× RF compression, while the total time (compression + reconstruction) was similar.

However, as in difference images, apparent artifacts appear at the bottom of the image given 4× RF compression, while 18× joint compression only incurred slight noise amplification relative to the non-compressed least-squares reconstruction and thus indicates higher compression efficiency.

Moreover, for compressed-sensing [34], joint compression cut reconstruction time to about 4.5 min only, while the peak memory required by 4× RF compression prevented CS reconstruction using the same reconstruction implementation (the reconstruction algorithm #2). The CS reconstruction based on joint compression achieved further improved image quality, close to the non-compressed, fully-sampled reference with identical field modulations.

As in Figure 8I–P, for linear-gradients-modulated scans, namely Wave-CAIPI/CS-Wave, least-squares reconstruction with 4 × 2 undersampling and compressed-sensing reconstruction with 10.5× undersampling are shown. For $L_2$ reconstructions, 16× joint compression incurred a long computational time given our current implementation. As a result, even though its reconstruction was 4.6× faster, the total time (compression + reconstruction) was 1.2× slower than the 4× RF compression. Note, the total computational time based on joint compression and $L_2$ reconstruction can also become less for other matrix sizes (e.g., 180 × 180 × 80 with 32 receivers in Figure S2, 230 × 230 × 144 with 44 receivers in Figure 9), compared to RF-only compression.

Nevertheless, the 16× joint compression leads to reduced artifacts, as indicated by difference maps. Moreover, it allowed compressed-sensing reconstruction (impossible with RF-only compression) in 4.58 min, achieving improved reconstruction quality compared with $L_2$, much closer to the non-compressed, fully-sampled reference with identical field modulations.

In Supporting Information (Figure S4), under identical compression factors (i.e., 8×, 16×, 32×), reconstructions using the RF-only compression show clear loss of image quality than the proposed joint compression techniques, for both 2D and 3D in vivo FLASH with local $B_0$ coil modulations at 9.4 T.

### 4.3 | RF-Only Versus $B_0$-RF Compression for 3 T In Vivo Scans

In Figure 9, reconstruction of 3 T Wave-CAIPI/CS-Wave in vivo scans is shown, comparing 2.2× RF-only compression (20 virtual RF coils) and 11× joint compression.

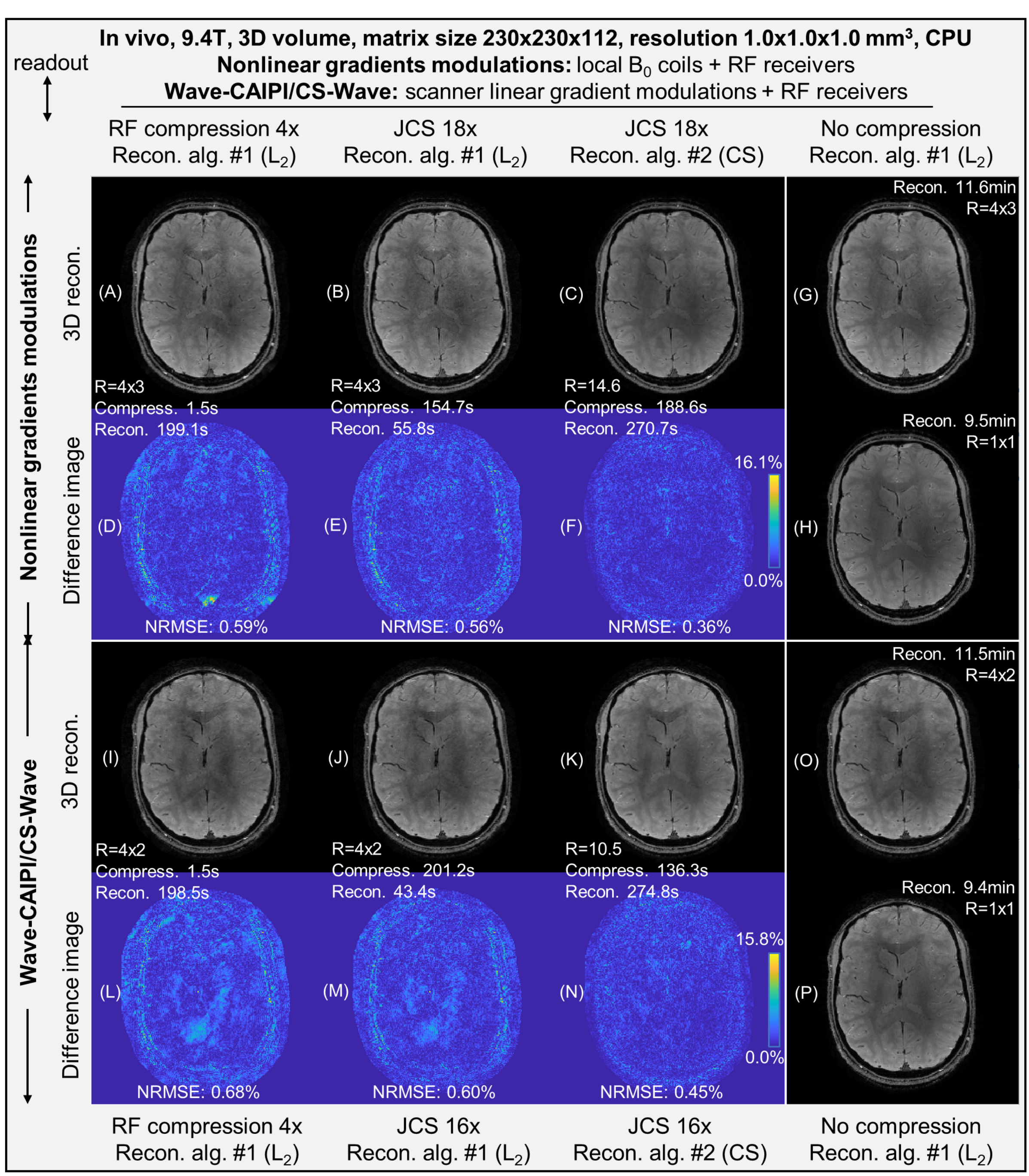

**FIGURE 8** | 3D least-squares and compressed-sensing reconstructions with reasonable compression factors for RF-only and joint compression algorithms at 9.4 T. The rapid field modulations were provided by local $B_0$ coils (A–H) and scanner's linear phase-encoding gradients (Wave-CAIPI $L_2$/CS-Wave, I-P). (A) $L_2$ reconstruction with 4× RF compression. (B) $L_2$ reconstruction with 18× joint compression. (C) CS reconstruction with 18× joint compression. (D–F) Corresponding difference maps relative to the non-compressed, fully-sampled reference. (G) $L_2$ reconstruction without compression. (H) Fully-sampled $L_2$ reconstruction without compression. (I–P) Similar organizations as in (A–H) but for Wave-CAIPI/CS-Wave. Note, (H) and (P) were also reconstructed from the same field-modulated (local $B_0$ or Wave-CAIPI) scans, and thus, the difference maps here cannot fully reflect field-calibration artifacts, nor subject motions. The compression and reconstruction time are shown. In $L_2$ reconstructions, MATLAB tool "parfor" was used.

For $L_2$ reconstructions with 3 × 3 undersampling, both moderate compression factors (empirically chosen for lower SNR at 3 T) yielded almost identical image quality compared to the non-compressed least-squares reconstruction. However, they reduced reconstruction time from 40.5 to 20.4 min (2.2× RF compression) and to 127.6 s (11× joint compression). For this dataset, the total time (compression + reconstruction) for joint compression is much less than that for RF compression.

For compressed-sensing with 9.2× PD-VD undersampling, reconstruction time was reduced to 10.1 min by joint compression, whereas 2.2× RF-only compression could not be run due to peak-memory limits using our reconstruction implementations. The CS reconstruction quality was further improved than $L_2$, and approached non-compressed, fully-sampled reference with the same linear gradients modulations.

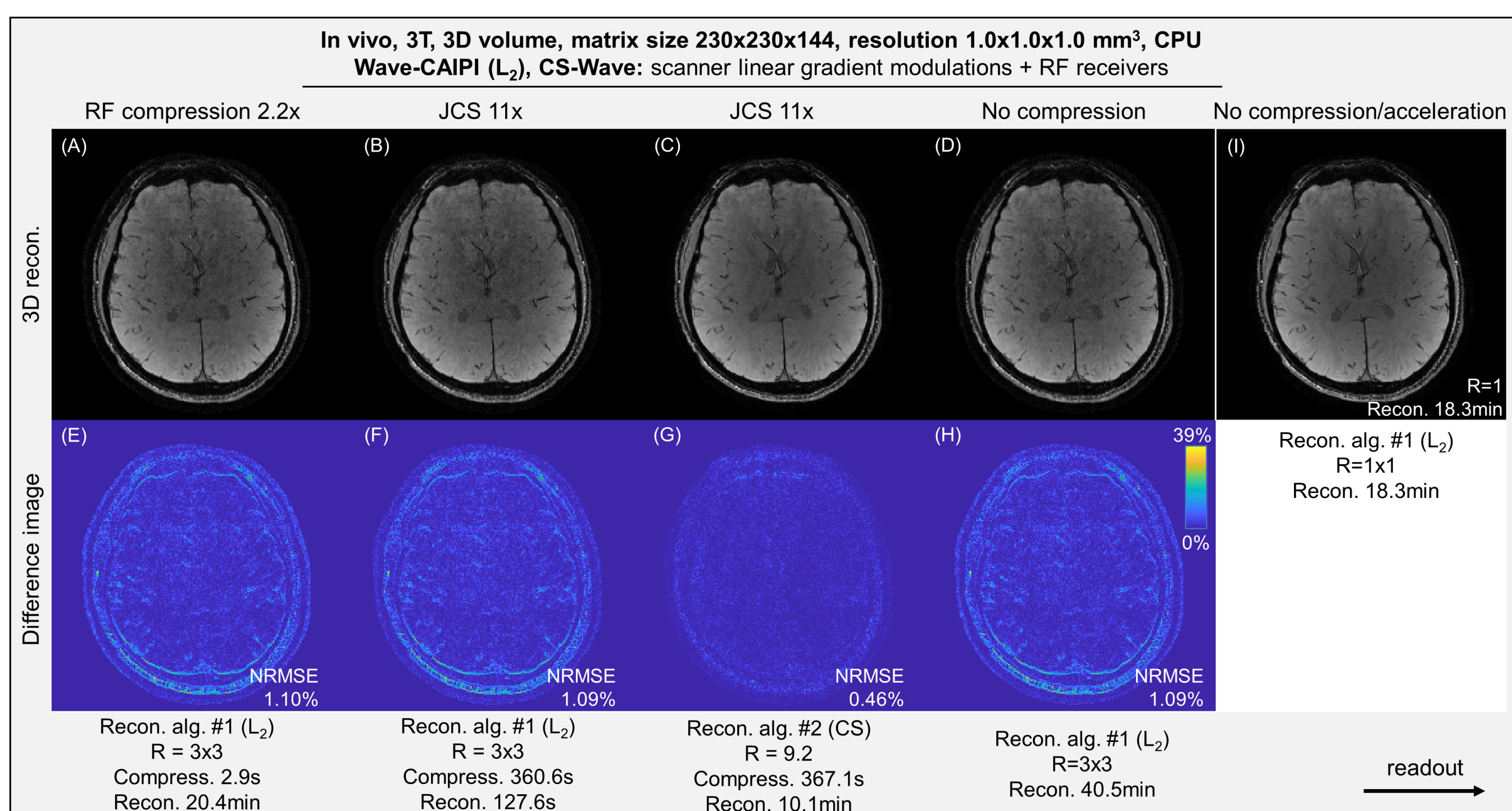

**FIGURE 9** | 3D least-squares and compressed-sensing Wave-CAIPI/CS-Wave reconstructions with reasonable compression factors for RF-only and joint compression algorithms at 3 T. (A) $L_2$ reconstruction using 2.2× RF compression with 3 × 3 undersampling. (B) $L_2$ reconstruction using 11× joint compression with 3 × 3 undersampling. (C) Compressed-sensing reconstruction using 11× joint compression with 9.2× PD-VD undersampling. (D) $L_2$ reconstruction without compression, with 3 × 3 undersampling. (E–H) Corresponding difference images relative to the non-compressed, fully-sampled least-squares reconstruction with the same field modulations. (I) Non-compressed, fully-sampled $L_2$ reconstruction of Wave-CAIPI. Note, (I) was also reconstructed from the same Wave-CAIPI scan, and thus, the difference maps here cannot fully reflect field-calibration artifacts, nor inter-scans subject motions. The compression and reconstruction time are shown. In $L_2$ reconstructions, MATLAB tool "parfor" was used.

### 4.4 | Visualization of Compressibility for In Vivo Scans at 9.4 T and 3 T

Figure 10 evaluates the singular-value distributions and compression thresholds obtained during RF array and joint compression algorithms for in vivo scans at 9.4 and 3 T (those in Figure S3, Figures 8 and 9, respectively), visualizing the signal subspaces on which the SVD operated, and their compressibility. They are calculated based on Equation (8).

RF-only compression discarded about 50%–75% of singular vectors in the receiver dimension (44 or 32 coils). In contrast, joint compression was far more aggressive, discarding 91%–95% of singular vectors in a larger joint space (5–16 oversampled readout points × 44/32 coils). Note that, although the RF sensitivity maps at 3 T could be generally more compressible than the maps with more complex standing-wave patterns at 9.4 T, we empirically applied less aggressive compression to the 3 T datasets. This is because of the lower intrinsic SNR at 3 T, which might make the encoding model more sensitive to over-compression.

## 5 | Discussion and Conclusions

We have demonstrated that 2D and 3D MRI scans accelerated by rapid $B_0$ field modulations and multiple RF receivers can be jointly compressed by about 11×–24×, dramatically exceeds the limits of conventional RF receiver array compression [38]. In our proof-of-concept MATLAB implementations on a high-memory CPU node in HPC, the proposed joint compression substantially reduced both reconstruction time and peak memory for 3D scans, enabling $L_2$ reconstruction in 19–127 s, CS reconstruction in 177 s-10.1 min, depending on the matrix-size and compression factors. Although 3D joint compression may require longer pre-processing time to compress data and maps (e.g., 47.2–367.1 s), this may start before scans finish and could be further optimized in future study (codes, GPU, programming languages [41, 68, 69]). By contrast, the CS reconstructions are largely hindered under RF-only compression, and, even at moderate compression factors, could suffer from degraded encoding efficiency compared to the ones using joint compression. For 2D scans, our proposed technique took 1.21–3.5 s for compression, enabling CS reconstruction within 1.4–5.1 s.

These improvements arise from the substantially reduced readout forward model, previously enlarged due to oversampled readout and multiple RF receivers—now becomes comparable in size to the data undersampling factors. The subregion-wise operation plays a crucial role: SVD acts for small, decoupled k-space patches, and the resulting set of compression matrices can be efficiently reused across other subregions sampled by the same subset of $B_0$ and RF kernels. Notably, our calculation steps to compress a composite spatial encoding model for a group is similar to the conventional algorithm to explicitly compress the conventional SENSE model [38], but our strategy to partition k-space in subregion groups and compress it in multidimensional patches is novel. "Pushing" these minimally compressible RF sensitivity maps into the highly redundant readout oversampled space with joint compression allows us to surpass the maximum achievable RF compression factor (usually 1.5×–3.0×) for the entire dataset. Therefore, the rapid $B_0$ modulations during oversampled readout eventually yield an information-enriched but meanwhile compact readout encoding model. Note, the significantly smaller forward model due to joint compression can theoretically make further GPU-acceleration much easier, as G-RAM is usually much more limited relative to CPU memory.

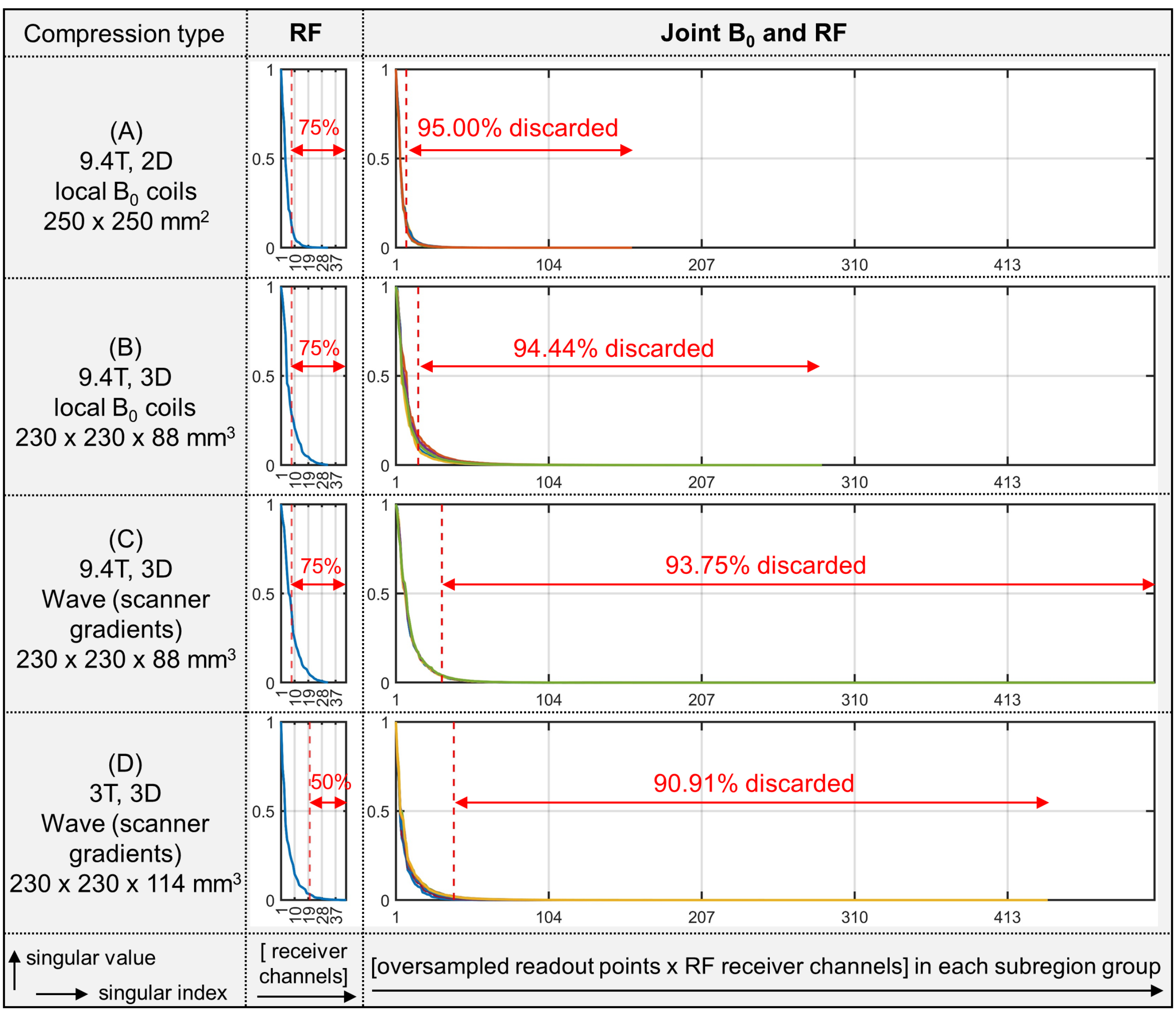

**FIGURE 10** | Singular-value distributions obtained in RF array compression and joint compressions for all in vivo scans at 9.4 T and 3 T in this paper. (A) 2D scans with local $B_0$ coils modulations at 9.4 T. (B) 3D scans with local $B_0$ coils modulations at 9.4 T. (C) 3D Wave-CAIPI/CS-Wave at 9.4 T. (D) 3D Wave-CAIPI/CS-Wave at 3 T.

Currently, in our proof-of-concept MATLAB scripts, the entire readout encoding matrix is explicitly stored in memory during compression and reconstruction, which is not the most memory-efficient implementation. Further optimization can mitigate this by exploiting the periodicity of the rapid $B_0$ modulation. Specifically, the readout encoding matrix only needs to be explicitly computed and stored for a single $B_0$ modulation period during compression, and repeatedly used to synthesize the entire readout duration during reconstruction in a memory-efficient manner.

In principle, our group-patch joint compression strategy can be beneficial when periodic $B_0$-RF encoding redundancy is present. For example, this can happen when rapid $B_0$ field modulations vary across TRs (e.g., readout CAIPI [30]), or contain multiple (e.g., 2.5–5–10 kHz) temporal frequencies along the readout time axis [29, 87]. As long as the entire k-space can still be partitioned into a small number of subregion groups that share identical $B_0$-RF modulation functions, the full data volume can be compressed efficiently based on a patch-by-patch manner in groups. While calculating and applying compressions may also be done entirely in k-space [39], or utilize neural networks [42], this paper's implementation yields virtual compressed maps in a straightforward manner to facilitate visual interpretations of the compression behaviors. Nevertheless, other compression algorithms might also be adapted for joint $B_0$-RF compression for performance comparisons in more details.

Theoretically, beyond reconstruction based on image-space to k-space mapping, the joint compression is also compatible with k-space interpolation-based reconstruction [88]. For a very high-resolution 3D dataset with rapid $B_0$ modulations and readout oversampling, joint compression can be applied prior to a generalized k-space reconstruction algorithm based on RKHS cardinal-function interpolations [47]. Such k-space reconstruction can be adapted in a similar subregion-wise manner [30, 80] and thus further decouples the segments along k-space readout dimension for faster reconstruction and lower-memory requirements. In the future, combining the subregion-wise joint compression and k-space interpolation-based reconstruction can be crucial; otherwise, 3D CS reconstruction may still incur high peak memory for high-resolution field-modulated scans required for certain applications [89].

### Acknowledgments

This study is supported by the ERC Advanced Grant No. 834940. The ex vivo brain phantom was courtesy of the Institute of Clinical Anatomy and Cell Analysis, Department of Anatomy, Eberhard Karls University of Tübingen. The first author thanks Dr. Thomas Shiozawa (Institute of Clinical Anatomy and Cell Analysis) for assistance with sample preparation, and Dr. Gisela Hagberg for assistance in scanning this phantom. The first author would like to thank Prof. Dr. Martin Uecker for discussions about compressed sensing reconstruction algorithms, Dr. Felix Breuer and Prof. Berkin Bilgic for helpful suggestions about CAIPIRINHA patterns in forward model reconstructions, Stefan Plappert for guidance in programming the ADwin high-speed processor, and Dr. Felix Glang, Dr. Jonas Bause, and Praveen Valsala for helpful comments about this manuscript. The first author thanks Dr. Franciszek Hennel and Prof. Dr. Klaas P. Pruessmann for interesting discussions about rapid magnetic field modulation techniques. Open Access funding enabled and organized by Projekt DEAL.

### Funding

This work was supported by the European Research Council (834940).

### Data Availability Statement

The example code or data that support the findings of this study is openly available in Zenodo at https://zenodo.org/records/20533698, in Github at https://github.com/ruitianwater/GroupPatchJCS_v1.0, and collected at the HarmonizedMRI website at https://harmonizedmri.github.io/projects/.

### Supporting Information

Additional supporting information can be found online in the Supporting Information section. **Figure S1:** The reconstruction of a 2D scan accelerated by rapid $B_0$ modulation and multiple RF receives, with no compression, RF array compression and joint compression. (A) k-space undersampling mask. (B) Reconstructed image. (C) Difference image between the reconstruction and the reference image without compression and undersampling. (D) SSIM comparing the reconstruction and the reference image without compression and undersampling. (E) computational profiling. With no undersampling and no compression, the reference image sets a baseline of 0.00% NRMSE and 1.00 SSIM. **Figure S2:** The reconstruction of a 3D scan accelerated by rapid $B_0$ modulation and multiple RF receives, with no compression, RF array compression and joint compression. (A) k-space undersampling mask. (B) Reconstructed image. (C) Difference image between the reconstruction and the reference image without compression and undersampling. (D) SSIM comparing the reconstruction and the reference image without compression and undersampling. (E) computational profiling. With no undersampling and no compression, the reference image sets a baseline of 0.00% NRMSE and 1.0 SSIM. **Figure S3:** 2D least-squares and compressed-sensing reconstructions with reasonable compression factors for RF-only and joint compression algorithms at 9.4 T. The rapid field modulations were provided by local $B_0$ coils. The Poisson-disc varaible-density (PD-VD) mask was generated by BART. (A–D): $L_2$ and CS reconstructions with 7× undersampling: 4× RF compression for $L_2$ reconstruction, 4× RF compression for compressed-sensing reconstruction, 20× joint compression for $L_2$ reconstruction, 20× joint compression for compressed-sensing reconstruction. (E–H): Different images relative to the non-compressed, fully-sampled image. (I) non-compressed, 7×-undersampled $L_2$ reconstruction. (J) non-compressed, fully-sampled $L_2$ reconstruction. Note, (J) was also reconstructed from the same local $B_0$ modulated scan, and thus, the difference maps here cannot fully reflect field-calibration artifacts, nor inter-scans subject motions. The compression and reconstruction time are shown. In $L_2$ reconstructions, MATLAB tool "parfor" was used. Here, the CS reconstruction incorporates additional wavelet and total variation regularizations, while still using a 1D regular undersampling mask for 2D reconstruction. **Figure S4:** 2D and 3D least-squares reconstructions with same compression factors at 9.4 T, comparing RF-only and joint compression algorithms. The rapid field modulations were provided by local $B_0$ coils. (A–D): 16× compression for a 2D scan with 7× undersampling. (E–H): 8× compression for a 2D scan with 7× undersampling. (I): reference fully-sampled 2D image without local $B_0$ modulations and compressions. (J–M): 32× compression for a 3D scan with 4 × 3 undersampling. (N–Q): 16× compression for a 3D scan with 4 × 3 undersampling. (R): reference fully-sampled 3D image without local $B_0$ modulations and compressions. Next to the reconstructions, the difference images were calculated relative to the reference FLASH images, which should be free of any artifacts caused by local $B_0$ coil or compression errors, but might contain subject motions relative to the field-modulated scans. The compression and reconstruction time are shown. Hybrid-space least-squares reconstruction with MATLAB tool "parfor" was used. **Figure S5:** Supporting Information for Figure S3: Accelerated data sampling enabled by the rapid $B_0$ modulations generated by the local $B_0$ coil array at 9.4 T. (A) The phase offsets in local $B_0$ coils, the magnetic field distribution generated at nearly-peak current, and the estimated gradient strength distribution (Euclidean-norm in two orthogonal dimensions) are shown. (B–E) The G-map, the G-factor boxplot for each of 12 slices, the k-space cardinal and power function maps are shown, which indicate sufficient sampling to reach 7× joint acceleration factor by $B_0$ modulations and SENSE. (F–K) The reference fully-sampled image with $B_0$ and SENSE reconstruction is successfully reconstructed. The $L_2$ reconstruction of 7× undersampled k-space has only subtle noise amplifications without noticeable residue fold-over artifacts. Its $L_1$ reconstruction further improves the image quality, as also shown in their difference images, and is very close to the reference image. In comparison, the SENSE-only acceleration has severe artifacts. Here, the CS $L_1$ reconstruction incorporates additional wavelet and total variation regularizations, while still using a 1D regular undersampling mask for 2D reconstruction. **Figure S6:** Supporting Information for Figure 8 of the manuscript: Accelerated data sampling enabled by the rapid $B_0$ modulations generated by the local $B_0$ coil array, or the scanner's phase-encoding gradients (i.e., Wave-CAIPI/CS-Wave), at 9.4 T. The k-space $B_0$ trajectory for local $B_0$ coils modulations is visualized using power function (effective k-space coverage) at a readout timepoint, considering a single RF receiver. The k-space $B_0$ trajectory for Wave-CAIPI is plotted from Pulseq built-in functions. The k-space power and cardinal function maps across three readout timepoints incorporate encoding by both rapid $B_0$ modulations and multiple RF receivers. The volumetric G-factors across the entire 3D volume are shown in the boxplots for the two different experiments, respectively. Note, the k-space efficiency maps based on the RKHS formalism are relative values and unitless. In both experiments, they show effective expansion of k-space coverage (thickened trajectories) by rapid $B_0$ modulations and multiple RF receivers. However, the detailed comparison between different experiments in the cardinal and power function maps are reserved for future study. **Figure S7:** Supporting Information for Figure 9 of the manuscript: Accelerated data sampling enabled by the rapid $B_0$ modulations generated by the scanner's phase-encoding gradients (i.e., Wave-CAIPI/CS-Wave) at 3 T. Left: least-squares reconstruction. Middle: G-map for the corresponding slice position. Right: G-factor boxplot across the entire 3D volume. **Table S1:** Summary of transitioning characteristics of different compression types and factors in Figure 7. **Table S2:** Computational profiling of different reconstruction algorithms. The dataset is a 3D GRE ex vivo scan accelerated by rapid $B_0$ field modulations and 32 RF receivers, given an image pixel size of $180 \times 180 \times 80$ and a k-space readout oversampling factor of 8. The "time" considers only the reconstruction time duration of the iterative algorithms, excluding the time for loading, saving and preparing the data and encoding matrix which were fed into the iterative reconstruction algorithms. Note the wavelet function in MATLAB was internally double-precision.

# Supporting Information for

# Group-Patch Joint Compression: Compressing Dynamic $B_0$ and Static RF Spatial Modulations Across k-Space Subregion Groups for Highly Accelerated MRI

Rui Tian[1]*, Klaus Scheffler[1,2]

[1]High-Field MR center, Max Planck Institute for Biological Cybernetics, Tübingen, Germany

[2]Department for Biomedical Magnetic Resonance, University of Tübingen, Tübingen, Germany

*Corresponding author: rui.tian@tuebingen.mpg.de

This Supporting Information provides additional technical details about the group-patch joint compression approach, and comparisons with the conventional RF receiver array compression. Technical information about the image acceleration by local $B_0$ coils-SENSE and Wave-CAIPI/CS-Wave are also shown.

### 1 Investigating data compression using ex-vivo phantom scans

In this section, more technical details of data compression and image reconstruction are presented, based on retrospectively undersampled ex-vivo phantom scans without in-vivo motions.

The Figure 7 of the manuscript can be summarized by the Table S1 as below, to help understanding:

| Compression | Encoding map | k-space "stamp" | Power function/TR | G-map |
|---|---|---|---|---|
| None | Physical sensitivity | GRAPPA kernel | Single trough | mean/max: 1.1/1.5 |
| 4x RF-only | Slightly altered | Slightly dispersed | Single trough | Worse |
| 4x joint | Strongly oscillatory | Extremely dispersed | Splits, Two troughs | Unchanged |
| 8x joint | Strongly oscillatory | Extremely dispersed | Two troughs, larger approximation errors | Unchanged |
| 12x joint | Oscillatory | Dispersed | Two closer troughs | Unchanged |
| 24x joint | Similar to physical $B_0$ and $B_1$ maps | Slightly dispersed | Slightly wider single trough | Slightly worse |
| 72x joint | Only most essential patterns retained | Small/compact | Narrower single trough, larger approximation errors | $L_2$ recon. impossible |

Table S1. Summary of transitioning characteristics of different compression types and factors in Figure 7.

The 2D and 3D ex-vivo (without motions) reconstructions using the reconstruction algorithm #2 (FISTA, compatible for random sampling masks) were investigated as shown in Figure S1-S2, which highlights variations in reconstruction quality and computational speeds (i.e., computing compression matrices, applying compression, reconstructing images) across different compression types and factors. This comparison excludes the reconstruction algorithm #1 (only compatible with regular/CAIPI sampling masks), focusing solely on the dependency of reconstruction performance on conventional RF-only compression versus our proposed joint compression.

### 1.1 2D compression, reconstruction algorithm #2

In Figure S1, a 2D FLASH ex-vivo scan accelerated by rapid $B_0$ modulations and multiple RF receivers was reconstructed with various compression factors for the readout forward model. We used a generic FISTA-based code (i.e., reconstruction algorithm #2) compatible with both least-squares (i.e., disabling nonlinear regularizations) and compressed sensing reconstruction to compare reconstruction speed and accuracy, with varying compression types and factors.

Figure S1A shows multiple sampling patterns, including fully sampling, 6x regular undersampling of phase-encoded steps, and the 6x undersampling pattern with additional 5 central k-space calibration lines (i.e., equivalent ACS lines with $B_0$ modulations).

The least-squares reconstruction with 6x-undersampling shows noise amplification as expected. Based on the visible image quality, and the quantitative metrics (i.e., NRMSE and SSIM), the comparison of regular 6x-undersampled reconstruction quality is (“>” means “better than”):

No compression > joint compression 24x > RF compression 4x

The computation time of joint compression is about 1.2s longer than RF compression, due to larger size of the encoding matrices, although this might start before the MRI scan finishes. However, the reconstruction time for the jointly compressed forward model is 3.4s less than the one by RF compression.

The compressed sensing reconstruction has even better reconstruction quality based on the NRMSE and SSIM, mainly due to nonlinear regularizations. The 24x joint compression makes the CPU reconstruction time within only 1.4s.

Overall, the joint-compression is much more efficient than the RF-only compression, although the 2D reconstruction with a pixel size of 180 x 180 doesn't cost that much time.

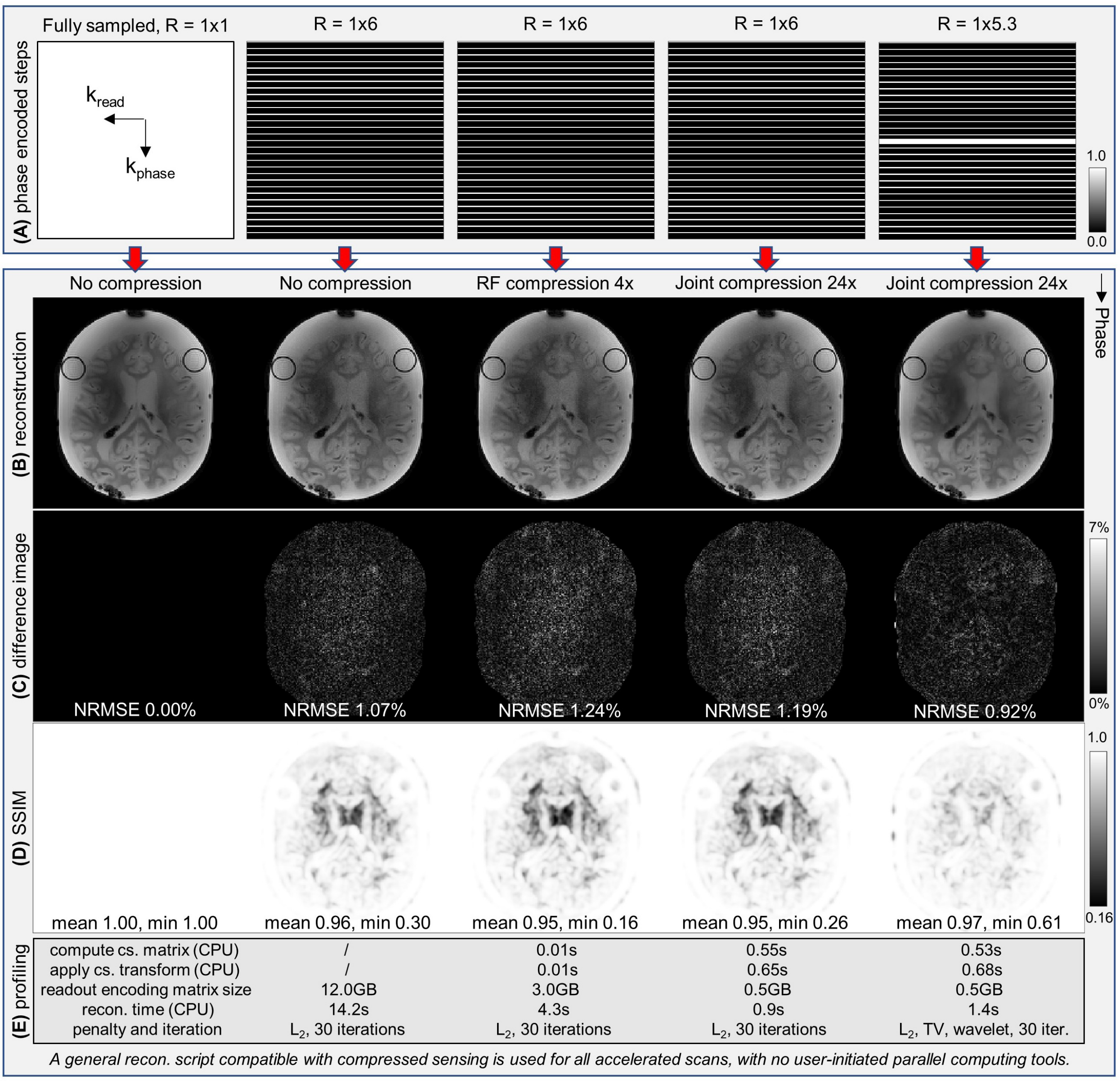

Figure S1. The reconstruction of a 2D scan accelerated by rapid $B_0$ modulation and multiple RF receives, with no compression, RF array compression and joint compression. A. k-space undersampling mask. B. reconstructed image. C.difference image between the reconstruction and the reference image without compression and undersampling. D. SSIM comparing the reconstruction and the reference image without compression and undersampling. E. computational profiling. With no undersampling and no compression, the reference image sets a baseline of 0.00% NRMSE and 1.00 SSIM.

**1.2 3D compression, reconstruction algorithm #2**

In Figure S2, a 3D FLASH ex-vivo scan accelerated by rapid $B_0$ modulations and multiple RF receivers was reconstructed with different compression factors for the readout forward model.

Figure S2A shows multiple 2D sampling patterns for the two phase-encoding dimensions, including fully sampling, regular undersampling using a 4x3 CAIPI pattern, VD-PD sampling patterns with a 10x10 fully sampled k-space center (i.e., equivalent ACS lines with rapid $B_0$ modulations).

The least-square reconstructions with 4x3 CAIPI undersampling exhibit noise amplifications as expected. However, based on the NRMSE and the SSIM, the reconstruction quality shows quantitative comparison, for both cases without and with nonlinear regularization (“>” means “better than”):

Joint compression 18x > RF compression 3.2x

The computational duration for jointly compressing larger matrices takes more time (here, ~47s) than RF only compression (~1s). However, this is negligible compared to the reconstruction time reduction due to jointly compressed readout forward model, dropping from 691.4s to 112.2s for least-square reconstruction, and decreased from 716.9s to 176.6s for compressed sensing reconstruction. This is because, correspondingly, the readout encoding matrix size was reduced from 298.6GB to 53.1GB.

Overall, the proposed joint compression for 3D scans is not only more efficient than RF array compression in terms of signal encoding, but also drastically lowers the computational demands. This ensures that the data expansion from highly-oversampled readout does not inappropriately exceed the undersampled data amounts, and thus, would not incur unreasonable computational burden especially for advanced iterative reconstruction using e.g., compressed sensing.

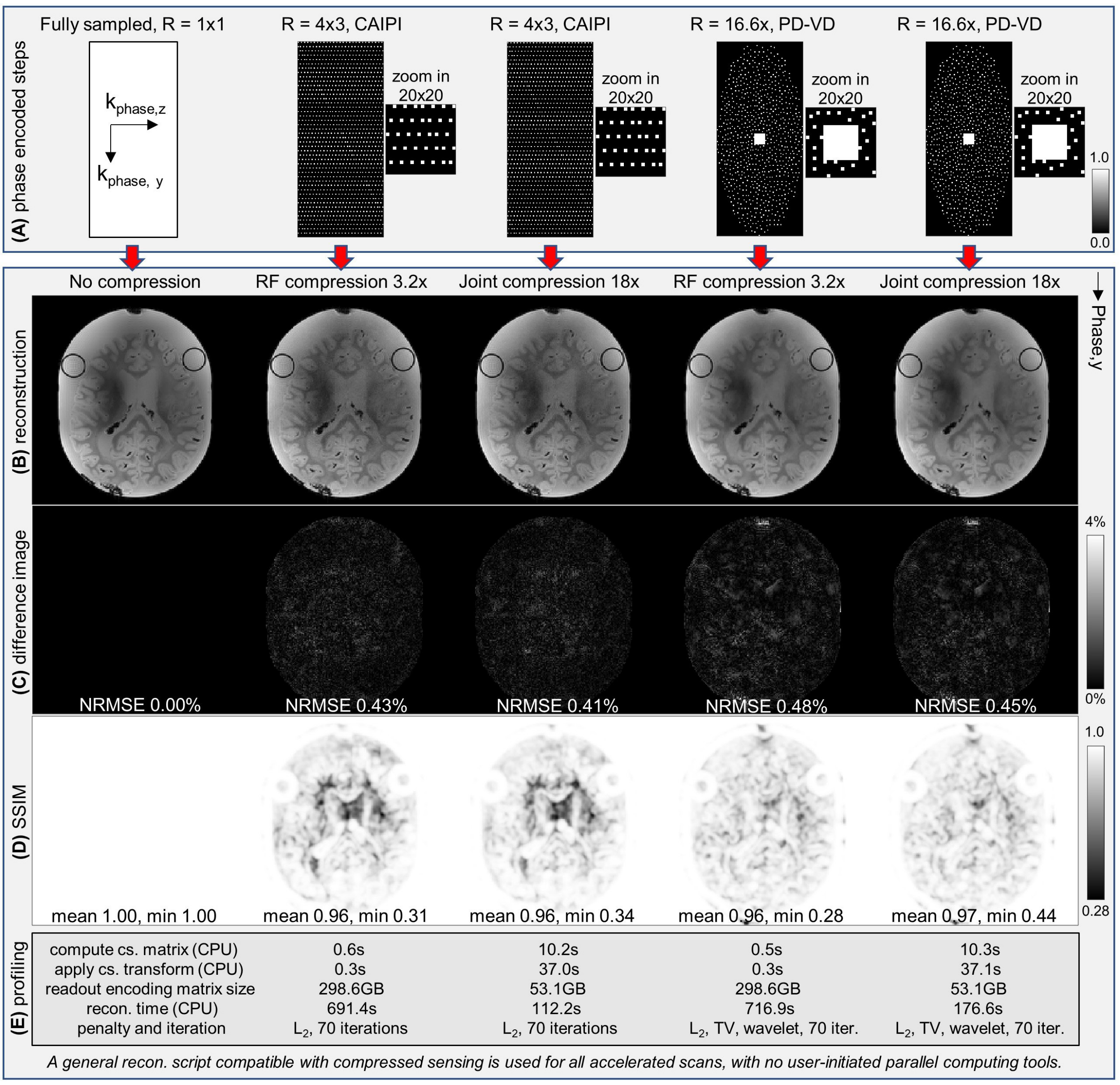

Figure S2. The reconstruction of a 3D scan accelerated by rapid $B_0$ modulation and multiple RF receives, with no compression, RF array compression and joint compression. A. k-space undersampling mask. B. reconstructed image. C.difference image between the reconstruction and the reference image without compression and undersampling. D. SSIM comparing the reconstruction and the reference image without compression and undersampling. E. computational profiling. With no undersampling and no compression, the reference image sets a baseline of 0.00% NRMSE and 1.0 SSIM.

## 1.3 3D compression, reconstruction algorithm #1 and #2

Furthermore, reconstruction times for the 3D ex-vivo FLASH scan with rapid $B_0$ modulations using both reconstruction algorithms (#1 and #2) are listed in the Table S2. This computational profiling helps guide the choice of reconstruction algorithms for different applications, such as varying sampling mask types or regularization strategies. In all cases of such matrix size, our joint compression technique substantially speeds up the reconstruction, compared to RF-only compression.

| **Hardware specification** | **The k-space undersampling pattern** | **3D recon. algorithms MATLAB, single pixel size 180x180x80 readout os. x8** | **Compress type** | **Read encoding matrix size** | **Time** | **Iteration** |
|---|---|---|---|---|---|---|
| HPC cluster<br>2xAMD<br>EPYC 7452<br>2.35GHz<br>64 cores<br>2 hyper-threads/core<br>1T MEM | R = 4x3<br>regular<br>CAIPI | read: decoupled, LSQR<br>phase: 2D-FFT<br>CPU | RF 3.2x | 0.2GB | 690s | auto |
| | | | Joint 18x | 0.04GB | 170s | auto |
| | | read: decoupled, LSQR<br>phase: 2D-FFT<br>CPU, “parfor” used | RF 3.2x | 0.2GB | 79s | auto |
| | | | Joint 18x | 0.04GB | 19s | auto |
| | | FISTA, $L_2$ only<br>read: full model<br>phase: 2D-FFT<br>CPU | RF 3.2x | 298.6GB | 691s | 70 |
| | | | Joint 18x | 53.1GB | 112s | 70 |
| | R = 16.6x<br>VD.<br>Poisson-Disc | FISTA, $L_2$+TV+wavelet<br>read: full model<br>phase: 2D-FFT<br>CPU | RF 3.2x | 298.6GB | 717s | 70 |
| | | | Joint 18x | 53.1GB | 177s | 70 |

Table S2. Computational profiling of different reconstruction algorithms. The dataset is a 3D GRE ex vivo scan accelerated by rapid $B_0$ field modulations and 32 RF receivers, given an image pixel size of $180 \times 180 \times 80$ and a k-space readout oversampling factor of 8. The “time” considers only the reconstruction time duration of the iterative algorithms, excluding the time for loading, saving and preparing the data and encoding matrix which were fed into the iterative reconstruction algorithms. Note the wavelet function in MATLAB was internally double-precision.

For decoupled phase-encoding pixels with regular 4x3 CAIPI undersampling pattern, the least-squares reconstruction was reduced from 690s using 3.2x RF compression, to 170s using 18x joint compression. Parallelizing the inversion of these decoupled readout forward models further reduced reconstruction time to 79s or 19s, respectively. For a generic script of FISTA (reconstruction algorithm #2) with an option to disable the nonlinear regularizations for comparing the least-square only computational time, the reduction factor of reconstruction time due to joint compression (112s) is similar to the reconstruction algorithm #1 that decouples sets of overlapped phase-encoding pixels without parallel CPUs (170s). The reconstruction algorithm #1 is not implemented with total variation and wavelet regularization. For compressed-sensing reconstruction with a Poisson-Disc undersampling mask, the FISTA took only 177s for reconstruction with jointly compressed readout forward model, substantially faster than the one based on RF array compression which took 717s.

In both cases, 18x joint compression delivered an additional 4x to 6x speed up over conventional RF-only compression, primarily because of a 5.6x smaller readout encoding matrix, without decreased encoding efficiency due to compression.

**2 RF-only vs. $B_0$-RF compression in-vivo 2D scans at 9.4T**

In Figure S3, with reasonable compression factors, reconstructions of 2D in-vivo FLASH with local $B_0$ coils modulations are shown, comparing the RF-only compression and the proposed group-patch joint compression.

Given 7x undersampling, the least-squares reconstruction (using the reconstruction algorithm #1) quality based on both 20x joint compression and 4x RF compression is nearly identical to the one without compressions. They have only slight noise amplifications relative to the non-compressed and fully-sampled least-squares reconstruction. With compressed-sensing reconstruction (using the reconstruction algorithm #2), the image quality based on both compression algorithms is nearly identical to the non-compressed, fully-sampled least-squares reconstruction. Here, the local $B_0$ modulations provided sampling incoherence for 2D compressed-sensing, where the 1D uniform undersampling pattern was used.

For joint compression, its compression time alone was slightly longer than RF compression, but its total time for compression and reconstruction became less, especially in compressed-sensing reconstructions (i.e., 3x faster).

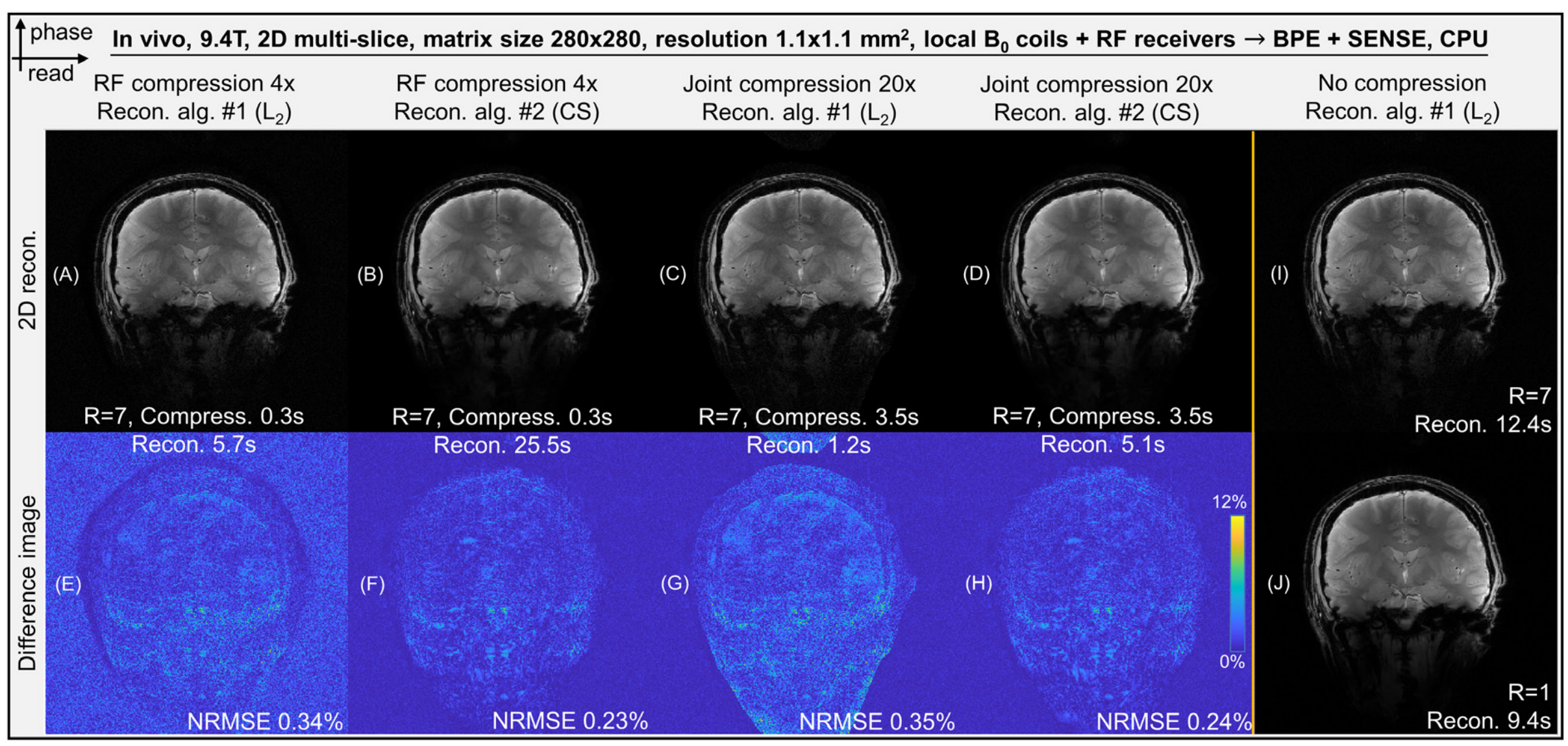

Figure S3. 2D least-squares and compressed-sensing reconstructions with reasonable compression factors for RF-only and joint compression algorithms at 9.4T. The rapid field modulations were provided by local $B_0$ coils. The varaible-density Poisson-disc (VD-PD) mask was generated by BART. (A-D): $L_2$ and CS reconstructions with 7x undersampling: 4x RF compression for $L_2$ reconstruction, 4x RF compression for compressed-sensing reconstruction, 20x joint compression for $L_2$ reconstruction, 20x joint compression for compressed-sensing reconstruction. (E-H): Different images relative to the non-compressed, fully-sampled image. (I) non-compressed, 7x-undersampled $L_2$ reconstruction. (J) non-compressed, fully-sampled $L_2$ reconstruction. Note, (J) was also reconstructed from the same local $B_0$ modulated scan, and thus, the difference maps here cannot fully reflect field-calibration artifacts, nor inter-scans subject motions. The compression and reconstruction time are shown. In $L_2$ reconstructions, MATLAB tool “parfor” was used. Here, the CS reconstruction incorporates additional wavelet and total variation regularizations, while still using a 1D regular undersampling mask for 2D reconstruction.

### 3 Comparing RF-only and joint compression under identical compression factors

In Figure S4, using identical compression factors, reconstructions using the RF-only and the proposed joint compression techniques are compared, for 2D and 3D in-vivo FLASH with local $B_0$ coil modulations at 9.4T.

In 2D FLASH with 7x undersampling, 16x RF compression (2 virtual coils) leads to strong artifacts. However, joint compression by both 16x and 8x have minimal artifacts. Note even moderate (8x) RF-only compression slightly degrades encoding efficiency than the joint compression, indicated by difference maps in the same scale.

In 3D FLASH with 4x3 undersampling, both 32x and 16x RF-only compression (1 and 2 virtual receiver coils) have strong artifacts. The 32x joint compression has only slight fold-over artifacts, which are removed in the 16x joint compression.

Note that, especially in 3D, joint compression takes noticeably longer than conventional RF array compression, because the SVD operates on a much larger signal space that additionally includes the oversampled readout points. Nevertheless, while compression can begin before the scan completes, the total computational time and peak memory for 2D and most 3D scans with joint compressions can still be substantially reduced compared to a realistic RF-only compression, as shown in manuscript.

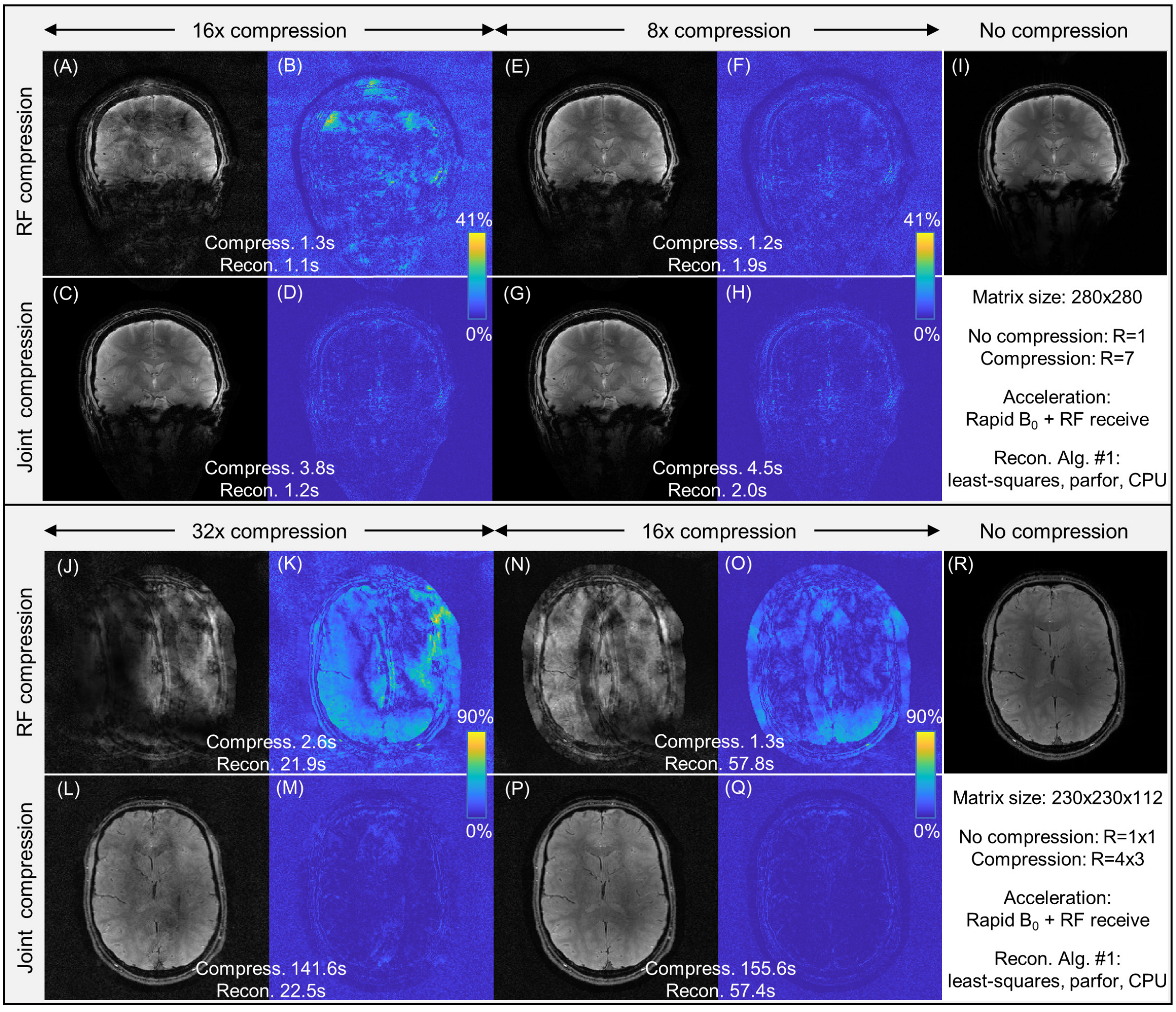

Figure S4. 2D and 3D least-squares reconstructions with same compression factors at 9.4 T, comparing RF-only and joint compression algorithms. The rapid field modulations were provided by local $B_0$ coils. (A–D): 16× compression for a 2D scan

with 7× undersampling. (E–H): 8× compression for a 2D scan with 7× undersampling. (I): reference fully-sampled 2D image without local $B_0$ modulations and compressions. (J–M): 32× compression for a 3D scan with 4 × 3 undersampling. (N–Q): 16× compression for a 3D scan with 4 × 3 undersampling. (R): reference fully-sampled 3D image without local $B_0$ modulations and compressions. Next to the reconstructions, the difference images were calculated relative to the reference FLASH images, which should be free of any artifacts caused by local $B_0$ coil or compression errors, but might contain subject motions relative to the field-modulated scans. The compression and reconstruction time are shown. Hybrid-space least-squares reconstruction with MATLAB tool "parfor" was used.

## 4 Increased encoding efficiency by rapid $B_0$ field modulations

This section provides supplementary information on encoding efficiency for in-vivo scans shown in Figure 8-9 of the manuscript, and Figure S3 in Supporting Information. Encoding efficiency for least-squares reconstruction can be accessed in either image-space or k-space. The image-space G-factor maps quantify pixel-wise noise amplification factors. The k-space cardinal function maps based on the RKHS formalism are the Frobenius norm of the k-space interpolation weights, and represent k-space noise amplification. The k-space power function maps based on the RKHS formalism characterize local k-space approximation error and represent the effective k-space sampling coverage. Here, we analyze encoding efficiency for the in-vivo scans in the manuscript, to highlight the acceleration potential in data acquisitions (note, not reconstruction speed) enabled by the rapid $B_0$ modulations, generated by either local $B_0$ coils or scanner's linear gradients. The encoding efficiency maps are computed from non-compressed encoding matrices. Note that, for G-map statistics, the boxplot in MATLAB is utilized. The central mark indicates the median, and the bottom and top edges of the box indicate the 25th and 75th percentiles, respectively. The whiskers extend to the most extreme data points not considered outliers, and the outliers are plotted individually using the red "+" marker symbol. These outlier G-map values can originate from those regions (e.g., background, skull) where $B_0$ field maps are difficult to calibrate, and thus, may not be meaningful to interpret the overall encoding efficiency. However, they are still plotted to remind the presence of these outliers.

As in Figure S5, for 2D coronal scans in Figure S3, we evaluate the $B_0$ field, the gradient field, the G-maps, the cardinal and power function maps, and compare reconstructions between SENSE-only acceleration and joint acceleration with rapid $B_0$ modulations and SENSE. The additional rapid $B_0$ modulations, when combined with multiple RF receivers, achieves a 7x acceleration factor for 2D FLASH, whereas conventional SENSE-only acceleration suffers from strong image artifacts.

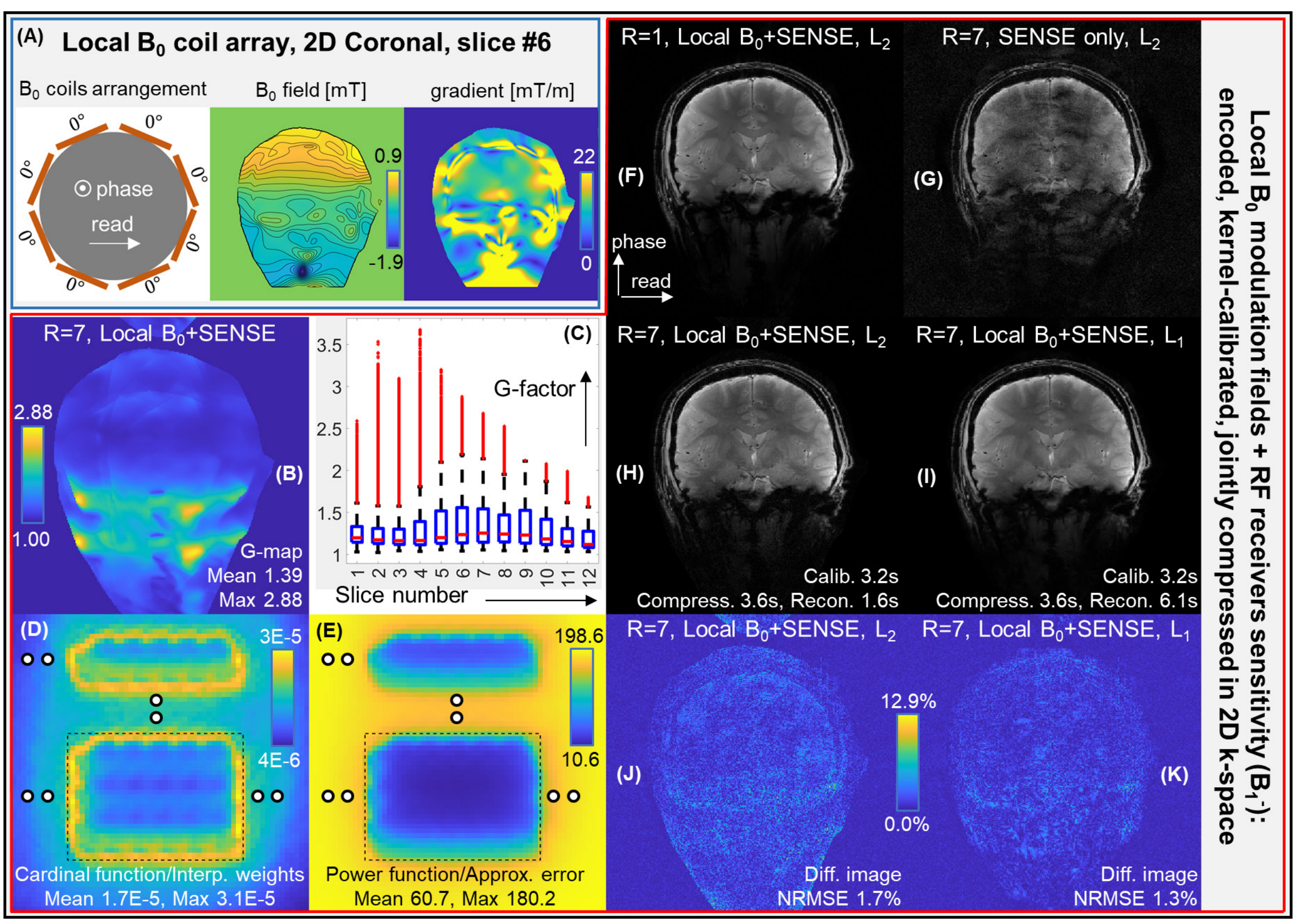

Figure S5. Supporting Information for Figure S3: Accelerated data sampling enabled by the rapid $B_0$ modulations generated by the local $B_0$ coil array at 9.4 T. (A) The phase offsets in local $B_0$ coils, the magnetic field distribution generated at nearly-peak current, and the estimated gradient strength distribution (Euclidean-norm in two orthogonal dimensions) are shown. (B–E) The G-map, the G-factor boxplot for each of 12 slices, the k-space cardinal and power function maps are shown, which indicate sufficient sampling to reach 7× joint acceleration factor by $B_0$ modulations and SENSE. (F–K) The reference fully-sampled image with $B_0$ and SENSE reconstruction is successfully reconstructed. The $L_2$ reconstruction of 7× undersampled k-space has only subtle noise amplifications without noticeable residue fold-over artifacts. Its $L_1$ reconstruction further improves the image quality, as also shown in their difference images, and is very close to the reference image. In comparison, the SENSE-only acceleration has severe artifacts. Here, the CS $L_1$ reconstruction incorporates additional wavelet and total variation regularizations, while still using a 1D regular undersampling mask for 2D reconstruction.

As in Figure S6, for 3D scans in Figure 8 of the manuscript, we show the k-space trajectory induced by rapid $B_0$ modulations, the k-space power function maps across readout timepoints, and the G-maps. Generally, the rapid $B_0$ modulations and multiple RF receivers yield reasonable encoding efficiency for such acceleration factors (i.e., 4x3, 4x2).

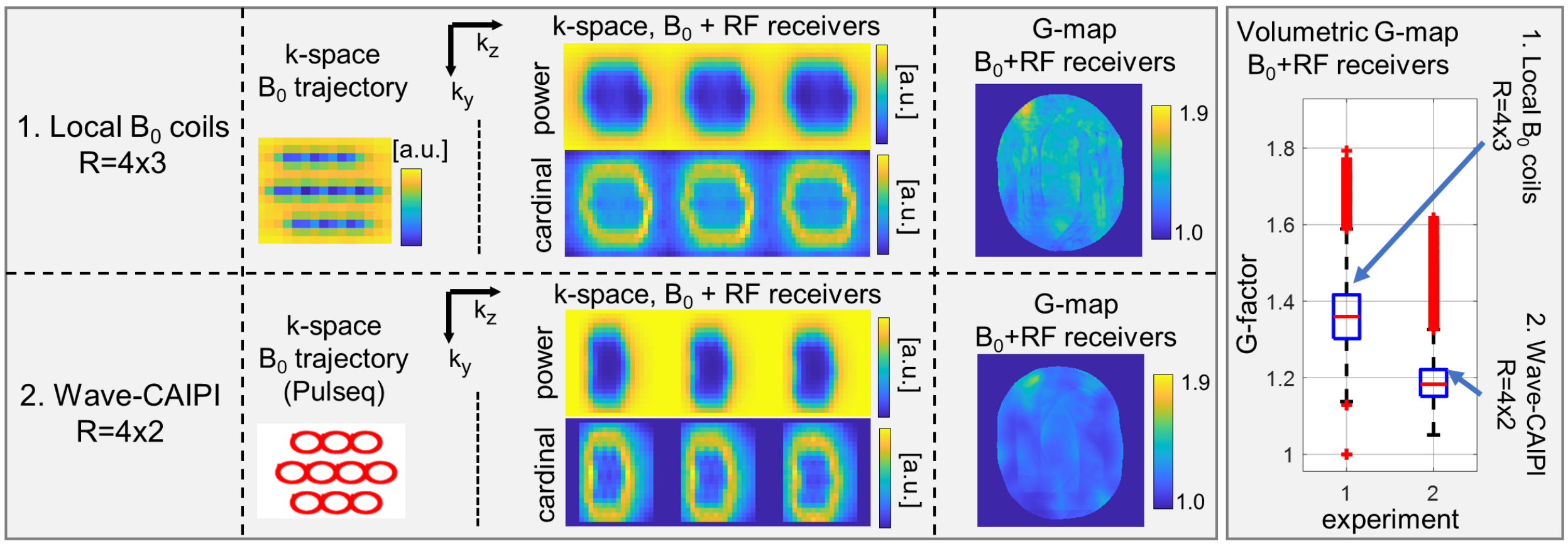

Figure S6. Supporting Information for Figure 8 of the manuscript: Accelerated data sampling enabled by the rapid $B_0$ modulations generated by the local $B_0$ coil array, or the scanner's phase-encoding gradients (i.e., Wave-CAIPI/CS-Wave), at 9.4 T. The k-space $B_0$ trajectory for local $B_0$ coils modulations is visualized using power function (effective k-space coverage) at a readout timepoint, considering a single RF receiver. The k-space $B_0$ trajectory for Wave-CAIPI is plotted from Pulseq built-in functions. The k-space power and cardinal function maps across three readout timepoints incorporate encoding by both rapid $B_0$ modulations and multiple RF receivers. The volumetric G-factors across the entire 3D volume are shown in the boxplots for the two different experiments, respectively. Note, the k-space efficiency maps based on the RKHS formalism are relative values and unitless. In both experiments, they show effective expansion of k-space coverage (thickened trajectories) by rapid $B_0$ modulations and multiple RF receivers. However, the detailed comparison between different experiments in the cardinal and power function maps are reserved for future study.

As in Figure S7, for 3D scans in Figure 9 of the manuscript, we show least-squares reconstruction, and the G-maps for the Wave-CAIPI scan. Here, the rapid $B_0$ modulations and multiple RF receivers provide reasonable encoding efficiency for the acceleration factor (i.e., 3x3) at 3T.

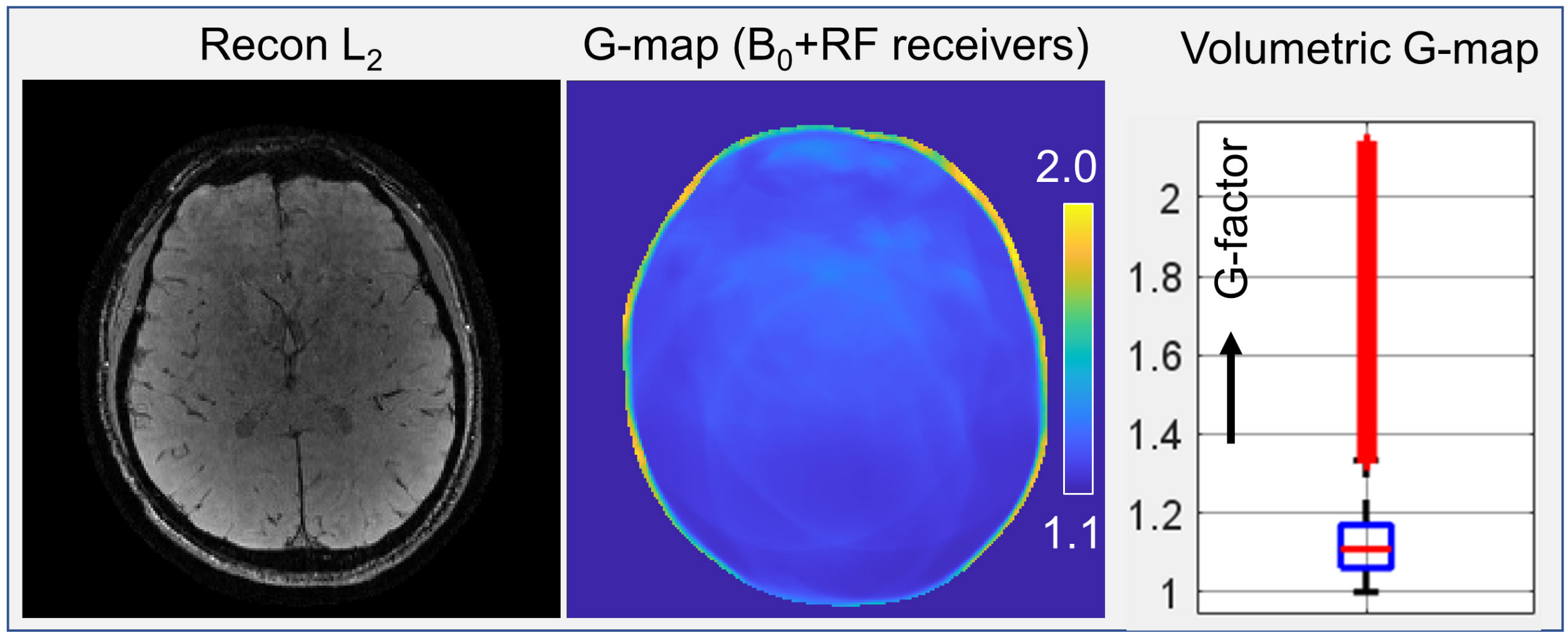

Figure S7. Supporting Information for Figure 9 of the manuscript: Accelerated data sampling enabled by the rapid $B_0$ modulations generated by the scanner's phase-encoding gradients (i.e., Wave-CAIPI/CS-Wave) at 3 T. Left: least-squares reconstruction. Middle: G-map for the corresponding slice position. Right: G-factor boxplot across the entire 3D volume.

**5 Sequence protocols**

In Figure 7, S1, the 2D multi-slice ex-vivo scan has a matrix-size of 180x180, corresponding to a resolution of 1.1mm x 1.1mm. The slice-thickness is 3 mm. The TE/TR is 4.91ms/30ms. The scanner's readout gradient bandwidth is 230Hz/Pixel. The 8-channel local $B_0$ coil array produced a nearly linear gradient with 7.41kHz/40$A_{pk}$ currents per channel, with one sinusoidal period spanning 45 oversampled data points.

In Figure S2, the 3D ex-vivo scan has a matrix-size of 180x180x80, which includes a 33.3% slice oversampling and corresponds to a resolution of 1.1mm x 1.1mm x 1.0mm. The TE/TR is 4 ms/30 ms. The scanner's readout gradient bandwidth is 230 Hz/Pixel. The local $B_0$ coil array produces a nearly linear gradient along the z phase-encoded direction, and a quadratic field along the transverse phase-encoded direction, oscillating in 7.41kHz/40$A_{pk}$ per channel, with one sinusoidal period spanning 45 oversampled data points.

In Figure S3, the 2D multi-slice in-vivo experiments was scanned with FOV 250x250$mm^2$, matrix-size 280x280, slice-thickness 3mm, readout-bandwidth 150Hz/Pixel, readout-duration 6.720ms, TE/TR 6.07ms/30ms. The $B_0$ coils currents were 7.41kHz/36$A_{pk}$ per channel. One sinusoidal period spanned 45 oversampled data points.

In Figure 8 and S4, the in-vivo 3D experiments with local $B_0$ coils modulations was scanned with FOV 230x230x88$mm^3$, matrix-size 230x230x112 (slice-oversampling 27.3%), readout-bandwidth 180Hz/Pixel, readout-duration 5.520ms, TE/TR 6ms/30ms. The $B_0$ coils currents were 7.41kHz/36$A_{pk}$ per channel. one sinusoidal period spanning 45 oversampled data points. In-vivo 3D Wave-CAIPI/CS-Wave (Pulseq) was designed similarly (i.e., identical FOV, matrix-size, readout-bandwidth, readout-duration, TE/TR. Nevertheless, the oscillating phase-encoding gradients (y-z) were (3.4-5.3) mT/m, 4.17kHz for wave-trajectory with 24 wave-cycles. One sinusoidal period spanned 80 oversampled data points.

In Figure 9, 3D Wave-CAIPI/CS-Wave (Pulseq) at 3T was scanned with 230x230x114mm$^3$, matrix-size 230x230x144 (slice-oversampling 26.3%), readout-bandwidth 180Hz/Pixel, readout-duration 5.520ms, TE/TR 20ms/35ms. The oscillating phase-encoding gradients (y-z) were (6.4-7.7) mT/m, 3.33kHz for wave-trajectory with 19 wave-cycles. One sinusoidal period spanned 100 oversampled data points.